\documentclass[a4paper]{article}

\usepackage{algorithm, algpseudocode, amsfonts, amsmath, amsopn, amssymb, amsthm, bbm, booktabs, color, enumitem, graphicx, multicol, multirow, rotating, tikz, url, xcolor}
\usepackage[a4paper]{geometry}

\newcommand{\real}{\mathbb{R}}
\newcommand{\zz}{\mathbb{Z}}

\newcommand{\one}{\mathbbm{1}}

\newcommand{\cL}{\mathcal{L}}
\newcommand{\cN}{\mathcal{N}}
\newcommand{\cO}{\mathcal{O}}
\newcommand{\cP}{\mathcal{P}}
\newcommand{\cX}{\mathcal{X}}

\newcommand{\myE}{\mathbb{E}}
\newcommand{\myP}{\mathbb{P}}

\newcommand{\myS}{{\rm S}}

\newcommand{\CRPS}{\textrm{CRPS}} 

\newcommand{\LogS}{\textrm{LogS}} 

\newcommand{\add}[1]{{\color{black}#1}}
\renewcommand{\it}{\color{blue}}

\newtheorem{assumption}{Assumption}
\newtheorem{theorem}{Theorem}

\begin{document}

\begin{center} 
		
\bf \Large 
		
Easy Uncertainty Quantification (EasyUQ): \\
Generating Predictive Distributions from \\ Single-valued Model Output
		
\medskip 

\normalsize
		
Eva-Maria Walz,\footnote{Institute for Stochastics, Karlsruhe Institute of Technology (KIT), Karlsruhe, Germany and Computational Statistics (CST) group, Heidelberg Institute for Theoretical Studies, Heidelberg, Germany (\url{eva-maria.walz@kit.edu})}   
Alexander Henzi,\footnote{Seminar for Statistics, ETH Zürich, Zürich, Switzerland (\url{alexander.henzi@stat.math.ethz.ch})} 
Johanna Ziegel,\footnote{Institute for Mathematical Statistics and Actuarial Science, University of Bern, Bern, Switzerland (\url{johanna.ziegel@stat.unibe.ch})}  
Tilmann Gneiting\footnote{Computational Statistics (CST) group, Heidelberg Institute for Theoretical Studies, Heidelberg, Germany and Institute for Stochastics, Karlsruhe Institute of Technology (KIT), Karlsruhe, Germany (\url{tilmann.gneiting@h-its.org}).}	
\bigskip
		
\today
		
\end{center}

\begin{abstract}
How can we quantify uncertainty if our favorite computational tool --- be it a numerical, a statistical, or a machine learning approach, or just any computer model --- provides single-valued output only?  In this article, we introduce the Easy Uncertainty Quantification (EasyUQ) technique, which transforms real-valued model output into calibrated statistical distributions, based solely on training data of model output--outcome pairs, without any need to access model input.  In its basic form, EasyUQ is a special case of the recently introduced Isotonic Distributional Regression (IDR) technique that leverages the pool-adjacent-violators algorithm for nonparametric isotonic regression.  EasyUQ yields discrete predictive distributions that are calibrated and optimal in finite samples, subject to stochastic monotonicity.  The workflow is fully automated, without any need for tuning.  The Smooth EasyUQ approach supplements IDR with kernel smoothing, to yield continuous predictive distributions that preserve key properties of the basic form, including both, stochastic monotonicity with respect to the original model output, and asymptotic consistency.  For the selection of kernel parameters, we introduce multiple one-fit grid search, a computationally much less demanding approximation to leave-one-out cross-validation.  We use simulation examples and forecast data from weather prediction to illustrate the techniques.  In a study of benchmark problems from machine learning, we show how EasyUQ and Smooth EasyUQ can be integrated into the workflow of neural network learning and hyperparameter tuning, and find EasyUQ to be competitive with conformal prediction, as well as more elaborate input-based approaches.
\end{abstract}

\section{Introduction}  \label{sec:introduction}

In an editorial that remains topical and relevant \cite{Trefethen2012}, SIAM President Nick Trefethen noted a decade ago that
\begin{quote} 
\small ``An answer that used to be a single number may now be a statistical distribution.''
\end{quote}
Indeed, with the increasing reliance of real world decisions on the output of computer models -- which might be numerical or statistical, parametric or nonparametric, simple or complex -- and the advent of uncertainty quantification as a scientific field of its own, there is a growing consensus in the computational sciences community that decisions ought to be informed by full predictive distributions, rather than single-valued model output.  For recent perspectives on these issues and uncertainty quantification in general, we refer to topical monographs \cite{Ghanem2017, Smith2014, Sullivan2015} and review articles \cite{Abdar2021, Berger2019, Gneiting2014, Roy2011}.

How can we quantify uncertainty if the computational model at hand provides single-valued output only?  With Nick Trefethen's comment in mind, we address the following problem: Given single-valued, univariate model output, how can we generate a prediction interval or, more generally, a probabilistic forecast in the form of a full statistical distribution?  In this work, we introduce the Easy Uncertainty Quantification (EasyUQ) technique that serves this task, based solely on a training archive of model output--outcome pairs.  The single-valued, univariate model output can be of any type --- e.g., it may stem from a physics-based numerical model, might arise from a purely statistical or machine learning model, or might be based on human expertise.  In a nutshell, EasyUQ applies the recently introduced Isotonic Distributional Regression (IDR, \cite{Henzi2021a}) approach to generate discrete, calibrated predictive distributions, conditional on the model output at hand.  The name stems from the three-fold reasons that EasyUQ operates on the final model output only, without any need for access to the original model input, that the method honors a natural assumption of isotonicity, namely, that higher values of the model output entail predictive distributions that are larger in stochastic order, and that the basic version of EasyUQ does not involve any tuning parameters, and thus does not require user intervention.  The more elaborate Smooth EasyUQ approach introduced in this paper subjects the EasyUQ distribution to kernel smoothing, to yield predictive probability densities that preserve key properties of the basic approach.  Prediction intervals are readily extracted; e.g., the equal-tailed 90\% interval forecast is framed by the quantiles at level 0.05 and 0.95 of the predictive distribution. 

As the EasyUQ approach requires training data, it addresses general ``weather-like'' tasks (\cite{Berger2019}, p.~441), which are characterized by frequent repetition of the task --- e.g., hourly, daily, monthly, at numerous spatial locations, or for a range of customers or patients --- in concert with short to moderate lead times of the forecasts, thus enabling the development of a sizeable archive of forecast--outcome pairs.  EasyUQ makes the best possible use of single-valued model output in the sense of empirical score minimization on the training data, subject to the natural constraint of isotonicity.  Specifically, the larger the model output, the larger the predictive distribution, in the technical sense of the familiar stochastic order \cite{Shaked2007}, i.e., the respective cumulative distribution functions (CDFs) do not intersect and their graphs move to the right as the model output increases.  Subject to the isotonicity constraint, the EasyUQ distributions are optimal with respect to a large class of loss functions that includes the popular continuous ranked probability score (CRPS, \cite{Gneiting2007a, Matheson1976}), all proper scoring rules for binary events, and all proper scoring rules for quantile forecasts, among others (\cite{Henzi2021a}, Thm.~2).  For prediction, the EasyUQ and Smooth EasyUQ distributions are interpolated to the value of the model output at hand, while respecting isotonicity.

\begin{figure}[p]
\centering
\includegraphics[width=0.9\textwidth]{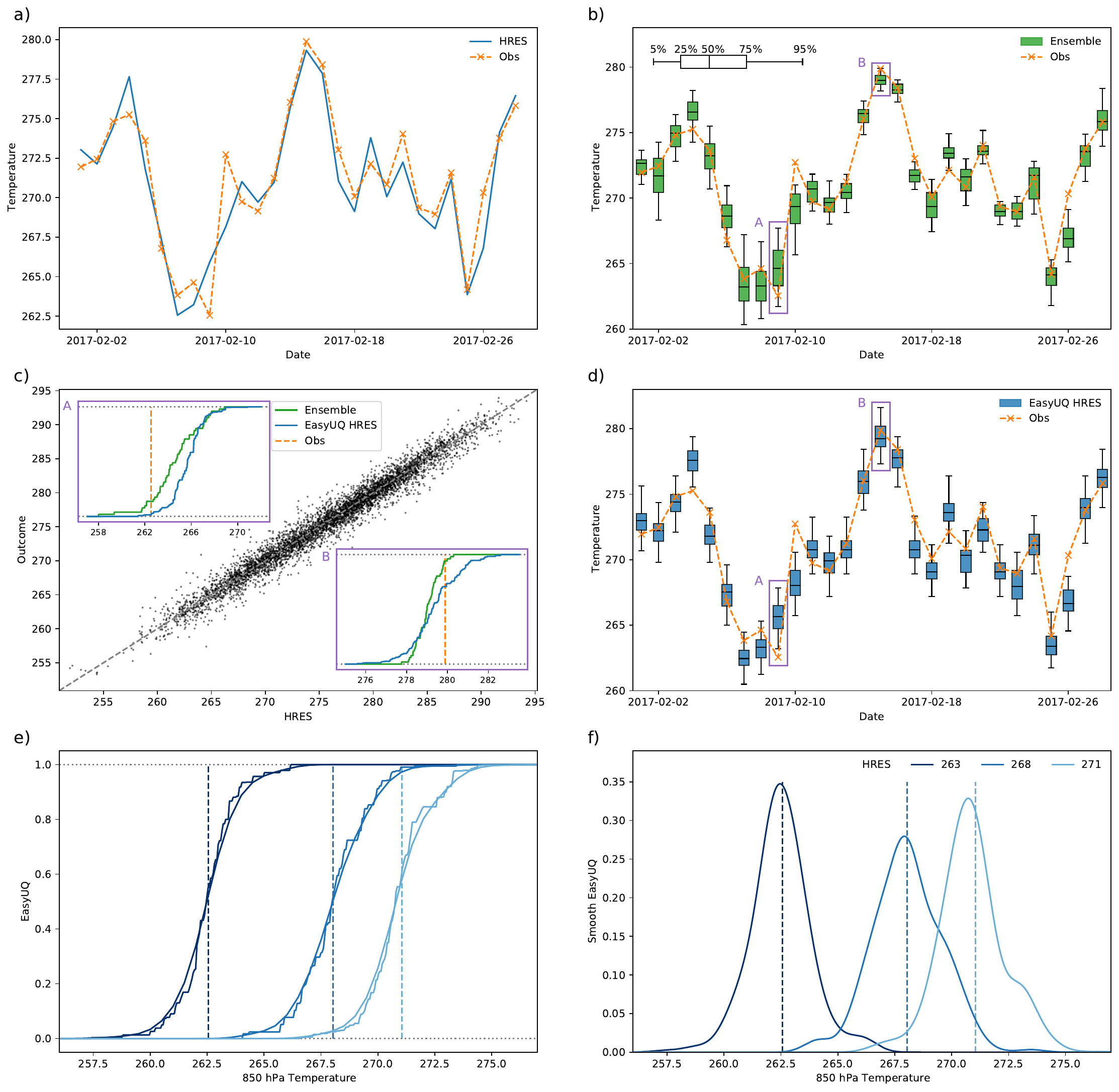}
\caption{EasyUQ illustrated on WeatherBench data.  Time series of three days ahead a) single-valued HRES model forecasts, b) state of the art ECMWF ensemble forecasts, and d) basic EasyUQ predictive distributions based on the single-valued HRES forecast along with associated outcomes of upper air temperature in February 2017 at a grid point over Poland, in degrees Kelvin.  The boxplots show the quantiles at level 0.05, 0.25, 0.50, 0.75, and 0.95 of the predictive distributions.  c) Scatterplot of HRES model output and associated outcomes in 2010 through 2016, which serve as training data.  The inset diagrams show the ECMWF and EasyUQ predictive CDFs for (A) 9 February 2017 and (B) 15 February 2017, respectively.  e) Basic and Smooth EasyUQ predictive CDFs and f) Smooth EasyUQ predictive densities at selected values of the single-valued HRES forecast.  For further details see Section \ref{sec:BasicWB}.  \label{fig:nutshell}}
\end{figure}

Figure~\ref{fig:nutshell} illustrates the EasyUQ approach on WeatherBench \cite{Rasp2020}, a benchmark dataset for weather prediction that serves as a running example in this paper.  Panel a) shows single-valued forecasts of upper air temperature from the HRES numerical weather prediction model run by the European Centre for Medium-Range Weather Forecasts (ECMWF, \cite{Molteni1996}) along with the associated observed temperatures in February 2017.  The training data for EasyUQ, which converts the single-valued HRES model output into conditional predictive distributions, comprise the forecast--outcome pairs from 2010 through 2016, as illustrated in the scatter plot in panel c).  Panel d) shows the EasyUQ predictive distributions for February 2017, which derive from the single-valued HRES forecasts in panel a), and can be compared to the computationally much more expensive ECMWF ensemble forecasts in panel b).  To facilitate the comparison, panel c) includes inset diagrams with the ECMWF ensemble and EasyUQ predictive CDFs for two particular days.  Panels e) and f) show EasyUQ predictive CDFs and Smooth EasyUQ predictive densities when the HRES model output equals 263, 268, and 273 degrees Kelvin, respectively.  The isotonicity property of the EasyUQ distributions is reflected by the non-intersecting CDFs.  The boxes in panels b) and d) range from the 25th to the 75th percentile of the distribution and generate 50\% prediction intervals, whereas the whiskers range from the 5th to the 95th percentile and form 90\% intervals. 

The remainder of the paper is organized as follows.  Section \ref{sec:Basic} provides comprehensive descriptions of IDR and the basic EasyUQ method, and we give detail, background information, and a comparison to conformal prediction \cite{Vovk2005, Vovk2020a} on both the WeatherBench temperature forecast challenge and a precipitation forecast example.  In Section \ref{sec:Smooth}, we introduce the Smooth EasyUQ technique, show that it retains the isotonocity property of the basic method, and discuss statistical large-sample consistency.  For the selection of kernel parameters, we introduce multiple one-fit grid search, a computationally much less demanding approximate version of cross-validation.  In Section \ref{sec:NN}, we demonstrate that EasyUQ can be integrated into the workflow of neural network learning and hyperparameter tuning, and use benchmark problems to compare its predictive performance to state-of-the-art techniques from machine learning and conformal prediction.  The paper closes with remarks in Section \ref{sec:discussion}, where we return to the discussion of input-based vs.~output-based uncertainty quantification.

While the basic version of EasyUQ arises as a special case of the extant IDR technique \cite{Henzi2021a}, we take the particular perspective of the conversion of single-valued model output into predictive distributions.  Original contributions in this paper include the development of the Smooth EasyUQ method (Sections \ref{sec:kernel} and \ref{sec:choice}), a detailed comparison to conformal prediction in case studies (Sections \ref{sec:BasicWB}, \ref{sec:BasicPrecip}, and \ref{sec:SmoothWBPrecip}) and from computational and methodological perspectives (Sections \ref{sec:computing} and \ref{sec:discussion}), and the integration and benchmarking of EasyUQ and Smooth EasyUQ for neural networks (Section \ref{sec:NN}).  In Appendix \ref{app:consistency}, we prove the consistency of smoothed CDFs in general settings, which supports the usage of Smooth EasyUQ, but is a result of broader and independent interest.

\section{Basic EasyUQ}  \label{sec:Basic}  

We begin the section with a prelude on the evaluation of predictions in the form of full statistical distributions.  Then we describe the IDR and EasyUQ techniques, and illustrate EasyUQ on the WeatherBench data from \cite{Rasp2020} and on precipitation forecasts \cite{Henzi2021a}.  Generally, EasyUQ depends on the availability of training data of the form
\begin{align}  \label{eq:data}
(x_i, y_i), \quad i = 1, \dots, n,
\end{align}
where $x_i \in \real$ is the single-valued model output and $y_i \in \real$ is the respective real-world outcome, for $i = 1, \ldots, n$.  For subsequent discussion, we note the contrast to more elaborate, input-based ways of uncertainty quantification that require access to the features or covariates from which the model output $x_i$ is generated.  In the WeatherBench example from Fig.~\ref{fig:nutshell}, we have training data comprising twice daily HRES forecasts and the associated observed temperatures in 2010 through 2016 as illustrated in panel c), where $n = 5,114$, but we do not have access to the excessively high-dimensional input to the HRES model.  In practice, one needs to find a predictive distribution given the value $x$ of the model output at hand, which may or may not be among the training values $x_1 \leq \cdots \leq x_n$, and some form of interpolation is needed, while retaining isotonicity.  In panel e) of Fig.~\ref{fig:nutshell} we illustrate predictive CDFs when $x$ equals equals 263, 268, and 273 degrees Kelvin, respectively.

Extensions of this setting to situations where single-valued output from multiple computational models is available can be handled within the IDR framework, as we discuss below.  If model output and real-world outcome are vector-valued --- e.g., when temperature is predicted at multiple sites simultaneously --- EasyUQ can be applied to each component independently, and the EasyUQ distributions for the components can be merged by exploiting dependence structures in the training data, based on empirical copula techniques such as the Schaake shuffle \cite{Schefzik2013}.

\subsection{Evaluating predictive distributions}  \label{sec:evaluating}

A widely accepted principle in the generation of predictive distributions is that sharpness ought to be maximized subject to calibration \cite{Gneiting2007b}.  Maximizing sharpness requires forecasters to provide informative, concentrated predictive distributions, and calibration posits that probabilities derived from these distributions conform with actual observed frequencies.  This is in line with and generalizes the classical goal of prediction intervals being as narrow as possible while attaining nominal coverage.

A key tool for evaluating and comparing predictive distributions under this principle are proper scoring rules \cite{Gneiting2007a, Matheson1976}, which are functions $\myS(P, y)$ mapping a predictive distribution $P$ and the outcome $y$ to a numerical score such that
\begin{align*}
\myE_{Y \sim P}[\myS(P, Y)] \leq \myE_{Y \sim P}[\myS(Q, Y)]
\end{align*}
for all distributions $P, Q$ in a given class $\cP$.  Here $\myE_{Y \sim P}[\cdot]$ denotes the expected value of the quantity in parentheses when $Y$ follows the distribution $P$.  From a decision-theoretic point of view, proper scoring rules encourage truthful forecasting, since forecasters minimize their expected score if they issue predictive distributions that correspond to their true beliefs.

Arguably the most widely used proper scoring rules are the continuous ranked probability score ($\CRPS$),
\begin{align}  \label{eq:CRPS}
\CRPS(F, y) = \int_{-\infty}^\infty (F(z) - \one\{z \geq y\})^2 \, \textrm{d}z,
\end{align}
which can be applied to cumulative distribution functions (CDFs) $F$ on the real line for which the corresponding distribution has finite first moment; and the logarithmic score for a predictive CDF $F$ with density $f$,
\begin{align}  \label{eq:LogS}
\LogS(F, y) = - \log(f(y)).
\end{align}
The popularity of the $\CRPS$ is due to the fact that it allows arbitrary types of predictive distributions (discrete, continuous, mixed discrete-continuous), is reported in the same unit as the outcomes, and reduces to the absolute error $\mathrm{AE}(x, y) = |x-y|$ if $F$ assigns probability one to a point $x \in \real$.  The $\LogS$ is (save for change of sign) the ubiquitous loss function in maximum likelihood estimation.  Closed form expressions for the $\CRPS$ and $\LogS$ are available for the most commonly used parametric distributions and have been implemented in software packages \cite{Jordan2019}.  In practice, forecast methods are compared in terms of their average score over a collection $(F_j, y_j)$ for $j = 1, \dots, n$, 
\begin{align*}
\bar\myS = \frac{1}{n} \sum_{j=1}^n \myS(F_j, y_j),
\end{align*}
and the method achieving the lowest average score is considered superior.

\subsection{Basic EasyUQ: Leveraging the Isotonic Distributional Regression (IDR) technique}  \label{sec:IDR}

In this section, it will be instructive to think of the quantities involved as random variables, which we emphasize by using upper case in the notation.  If model output $X$ serves to predict a future quantity $Y$, then one typically assumes that $Y$ tends to attain higher values as $X$ increases; in fact, the isotonicity assumption can be regarded as a natural requirement for $X$ to be a useful forecast for $Y$.  Isotonic Distributional Regression (IDR) is a recently introduced, nonparametric method for estimating the conditional distributions of a real-valued outcome $Y$ given a covariate or feature vector $X$ from a partially ordered space under general assumptions of isotonicity \cite{Henzi2021a}.  EasyUQ leverages the basic special case of IDR where $X$ is the single-valued model output at hand.  We review the construction and the most relevant properties of IDR for uncertainty quantification; for detailed formulations and proofs we refer to the original paper.

Formally, EasyUQ assumes that the conditional distributions of the outcome $Y$ given the model output $X$, which we identify with the CDFs $F_x(y) = \myP(Y \leq y \mid X = x)$, are increasing in stochastic order \cite{Shaked2007} in $x$, i.e., $F_x(y) \geq F_{x'}(y)$ for all $y \in \real$ if $x \leq x'$, or equivalently $q_x(\alpha) \leq q_{x'}(\alpha)$ for all $\alpha \in (0,1)$, where $q_x(\alpha) = F_x^{-1}(\alpha)$ is the conditional lower $\alpha$-quantile.  In plain words, the probability of the outcome $Y$ exceeding any threshold $y$ increases with the model output $x$.  Isotonicity in this sense is a natural assumption that one expects to hold, to a reasonable degree of approximation, in many types of applications.  An important exception arises under location-scale families.  Specifically, the arguments in the proof of Proposition 1 in Gneiting and Vogel \cite{Gneiting2022} imply that isotonicity is violated when the true predictive distributions come from a location-scale family with varying scale.\footnote{For example, if $F_1 = \cL(Y | X = x_1) = \cN(\mu_1, \sigma_1^2)$ and $F_2 = \cL(Y | X = x_2) = \cN(\mu_2, \sigma_2^2)$, where $x_1 \not= x_2$ and $\sigma_1 \not= \sigma_2$, then $F_1$ and $F_2$ are incomparable in stochastic order, whence isotonicity is violated.  However, if $\sigma_1$ and $\sigma_2$ are close to each other, the CDFs of $F_1$ and $F_2$ cross in the far (left or right) tail only (\cite{Gneiting2022}, proof of Proposition 1), so violations remain minor.}  However, the practical impact of this result is limited, due to the fact that in typical practice the scale parameter varies mildly only \cite{Gneiting2005b} and violations remain minor.  Crucially, estimators that enforce isotonicity tend to be superior to estimators that do not, even when the key assumption is violated, provided the deviation from isotonicity remains modest.  For an illustration in a simulation setting see the non-isotonic scenario (25) in Table 1 of Henzi et al.~\cite{Henzi2021a}, where IDR retains acceptable performance relative to its competitors, despite the key assumption being violated.  For a rigorous result, Thm.~7 of El Barmi and Mukerjee \cite{ElBarmi2005} demonstrates that, in the special case of discrete model output, EasyUQ has smaller large sample estimation error than non-isotonic alternatives even under mild violations of the isotonicity assumption.

EasyUQ assumes isotonicity with respect to the usual stochastic order.  In situations where this assumption is severely violated it may be worthwhile to consider isotonicity with respect to a weaker requirement for distributions to be ordered.  An analogous method to IDR under increasing concave and convex stochastic ordering constraints has been introduced by \cite{Henzi2022}.  An extension of EasyUQ in this direction is left for future work.

To estimate conditional CDFs under the given stochastic order constraints from training data of the form \eqref{eq:data}, we define
\begin{align}  \label{eq:IDR}
(\hat{F}_{x_1}(y), \dots, \hat{F}_{x_n}(y))' = \arg \min_{\theta \in \real^n \, : \, \theta_i \geq \theta_j \text{ if } x_i \leq x_j} \ \sum_{i = 1}^n (\theta_i - \one\{y_i \leq y\})^2
\end{align}
at $y \in \real$.  If $x_1 < \cdots < x_n$, then by classical results about isotonic regression,
\begin{align}  \label{eq:hat_F}
\hat{F}_{x_j}(y) = \min_{k = 1,\dots,j} \max_{l = j,\dots,n} \ \frac{1}{l-k+1} \sum_{i = k}^l \one\{ y_i \leq y \}, \quad j = 1, \ldots, n.
\end{align}
At any single threshold $y$, the computation can be performed efficiently in $\cO(n\log(n))$ complexity with the well-known pool-adjacent-violators (PAV) algorithm.  Since the loss function in \eqref{eq:IDR} is constant for $y$ in between the unique values $\tilde{y}_1 < \cdots < \tilde{y}_k$ of $y_1, \dots, y_n$, is suffices to compute \eqref{eq:hat_F} at the unique values, for which efficient recursive algorithms are available \cite{Henzi2022}.  An estimate $\hat{F}_x$ for the conditional CDF at model output $x \in (x_i, x_{i+1})$ is obtained by pointwise linear interpolation in $x$.  For $x \leq x_1$ and $x \geq x_n$, we use $\hat{F}_{x_1}$ and $\hat{F}_{x_n}$, respectively.  The EasyUQ conditional CDFs are step functions that correspond to discrete predictive distributions with mass at (a subset of) the unique values $\tilde{y}_1 < \cdots < \tilde{y}_k$ only.  

The IDR approach has desirable properties that make it suitable for uncertainty quantification.  By \eqref{eq:IDR}, the EasyUQ CDFs depend on the order of $x_1, \dots, x_n$ only, but not on their values, hence the solution is invariant under strictly monotone transformations of the model output, except for interpolation choices when $x \not\in \{x_1, \dots, x_n\}$.  Furthermore, the EasyUQ distributions are in-sample calibrated (\cite{Henzi2021a}, Thm.~2).  Importantly, a comparison of the loss function in \eqref{eq:IDR} and the definition of the $\CRPS$ in \eqref{eq:CRPS} reveals that EasyUQ minimizes the $\CRPS$ over all conditional distributions satisfying the stochastic order constraints.  Furthermore, the EasyUQ solution is universal, in the sense that it is simultaneously in-sample optimal with respect to comprehensive classes of proper scoring rules in terms of conditional CDFs or conditional quantiles, such as, e.g., weighted forms of the $\CRPS$ with the Lebesgue measure in \eqref{eq:CRPS} replaced by a general measure (\cite{Henzi2021a}, Thm.~2).  Other approaches to estimating conditional CDFs, e.g., based on parametric models, nearest neighbors, or kernel regression, do not share the universality property, and estimates change depending on the loss function at hand.

\begin{figure}[p]
\centering
\includegraphics[width=0.9\textwidth]{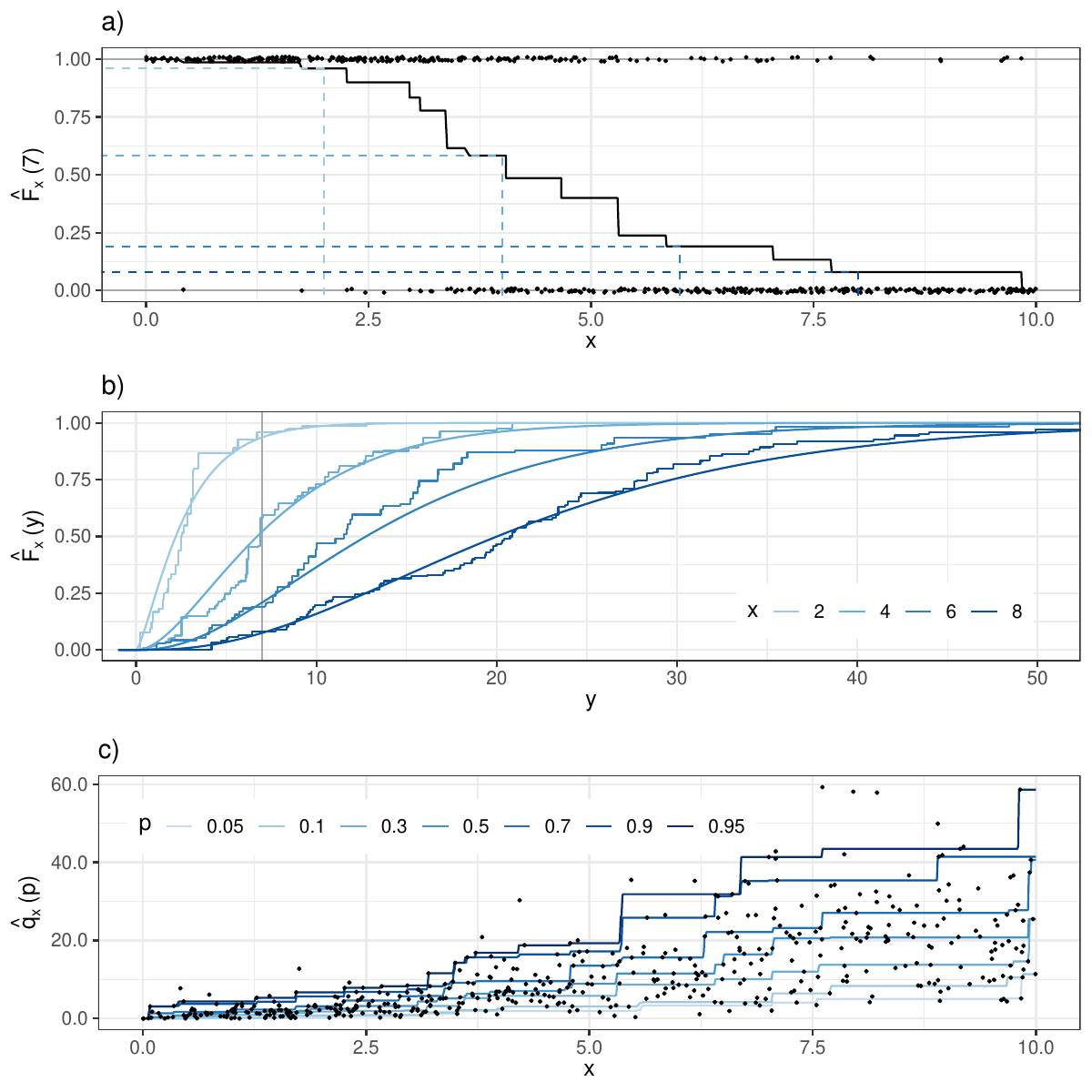}
\caption{Computation of EasyUQ predictive distributions from a training archive of $n$ = 500 model output--outcome pairs simulated according to \eqref{eq:sim}.  a) The minimizer $\hat{F}_x(y)$ of \eqref{eq:hat_F} at $y$ = 7, interpolated linearly in $x$.  The jiggled dots show the indicators $\one\{y_i \leq y\}$.  b) EasyUQ conditional CDFs $\hat{F}_x$ (step functions) and the respective true conditional CDFs (smooth curves) at selected values of $x$.  The vertical line at $y$ = 7 highlights the values marked in the top panel.  c) Training data $(x_i, y_i)$ for $i$ = 1, \dots, n, and conditional quantile curves $\hat{q}_x(p)$ resulting from inversion of the EasyUQ CDFs $\hat{F}_x$.  The lowest and highest quantile curves (levels 0.05 and 0.95) together delineate equal-tailed 90\% prediction intervals.  \label{fig:EasyUQ}}
\end{figure}

In Fig.~\ref{fig:nutshell} we illustrate EasyUQ predictive CDFs in the empirical WeatherBench example.  Simulation examples, to which we turn now, have the advantage of the true conditional CDFs being available, so we can compare to them.  Figure~\ref{fig:EasyUQ} illustrates the construction of the discrete EasyUQ predictive distributions step by step, based on a training archive of the form \eqref{eq:data} with $n = 500$ simulated from a bivariate distribution, where the model output $X$ is uniform on $(0,10)$ and the outcome $Y$ satisfies
\begin{align}  \label{eq:sim}
Y \mid X \sim \text{Gamma}(\text{shape} = \sqrt{X}, \text{scale} = \min\{ \max \{ X, 2\} , 8\}).
\end{align}
EasyUQ converts the single-valued model output $X$ into conditional predictive CDFs close to the right-skewed true ones.  Indeed, IDR, and hence, EasyUQ are asymptotically consistent: As the training archive size $n$ grows, the estimated EasyUQ CDFs converge to the true conditional CDFs \cite{ElBarmi2005, Henzi2021a, Moesching2020}.  Of particular relevance to EasyUQ is the following recent result (\cite{Henzi2021b}, Thm.~5.1): If $x_1, \dots, x_n$ themselves are not fixed but predictions from a statistical model that is estimated on the same training data then IDR is a consistent estimator of the true conditional distributions, subject to mild regularity conditions.

The basic EasyUQ method extends readily to vector-valued model output.  If $x_1, \dots, x_n$ are vectors in a space with a partial order $\preceq$, then the same approach \eqref{eq:IDR} applies with the usual inequality $\leq$ replaced by the partial order $\preceq$.  This allows more flexibility in the sense that distributions $F_x$ and $F_{x'}$ are allowed to be incomparable in stochastic order if $x$ and $x'$ are incomparable in the partial order.  A prominent example concerns ensemble weather forecasts \cite{Gneiting2005a, Leutbecher2008, Palmer2000}, where a numerical model is run several times under distinct conditions, and the partial order $\preceq$ that underlies IDR can be tailored to this setting \cite{Henzi2021a}.

To summarize, the basic EasyUQ method provides a data driven, theoretically principled, and fully automated approach to uncertainty quantification that is devoid of any need for implementation choices.  Based on training data, EasyUQ converts single-valued model output into calibrated predictive distributions that reflect the uncertainty in the model output and training data, as opposed to tuning intense methods, where uncertainty quantification might reflect implementation decisions and user choices.  The EasyUQ predictive solution is invariant under strictly monotone transformations of the model output, it is in-sample calibrated, it is in-sample optimal with respect to comprehensive classes of loss functions, and subject to mild conditions it is asymptotically consistent both for output from deterministic models, and output from statistical or machine learning models, even when the model is learned on the same data.\footnote{By Thm.~2 of Henzi et al.~\cite{Henzi2021a}, the fitted EasyUQ distributions are threshold calibrated, i.e., the predicted non-exceedance probabilities equal their empirical counterparts in the training data.  Furthermore, the fitted distributions are empirical score minimizers under a large class of proper scoring rules.  For discussion of the regularity conditions for asymptotic consistency, we refer to Appendix \ref{app:consistency} in this paper, Section 5 in \cite{Henzi2021b} and Section 2.4 in \cite{Henzi2021a}.}

\subsection{Illustration on WeatherBench challenge}  \label{sec:BasicWB} 

In a notable development, WeatherBench \cite{Rasp2020} introduces a benchmark dataset for the comparison of purely data driven and numerical weather prediction (NWP) model based approaches to weather forecasting.  Following up on the illustration in Fig.~\ref{fig:nutshell}, where we consider a grid point at (latitude, longitude) values of (53.4375, 16.875), we now provide background information and quantitative results at grid points worldwide.  

Our experiments are based on the setup in WeatherBench and consider forecasts of upper air temperature at a vertical level of 850 hPa pressure.  The forecasts are issued twice daily at 00 and 12 Coordinated Universal Time (UTC) at lead times of three and five days ahead.  The single-valued HRES forecast is from the high-resolution model operated by the European Centre for Medium-Range Weather Forecasts (ECMWF), which represents the physics and chemistry of the atmosphere and is generally considered the leading global NWP model.  To reduce the amount of data, WeatherBench regrids the HRES model output and the respective outcomes, which originally are on a 0.25 degree latitude--longitude grid ($72 \times 144$), to coarser resolution ($32 \times 64$) via bilinear interpolation.  The CNN forecast also is single-valued; it is purely data driven and based on a Convolutional Neural Network (CNN), with trained weights being available in WeatherBench.  The single-valued Climatology forecast is the best performing baseline model from WeatherBench; it is obtained as the arithmetic mean of the observed upper air temperature in the training data, stratified by 52 calender weeks.

Conformal Prediction (CP, \cite{Vovk2005, Vovk2020a}) is an increasingly popular, general technique for the construction of predictive distributions from single-valued model output.  For a comparison with EasyUQ, we employ CP in the form of the studentized Least Squares Prediction Machine (LSPM, \cite{Vovk2005}, Algorithm 7.2) with the single-valued model output as sole covariate.  We consider CP to be a key competitor, as it is an output-based method that shares desirable properties of EasyUQ.  Specifically, the LSPM supplements a least squares based point prediction of the outcome with a conformal predictive system for uncertainty quantification.  Based on training data $(x_i, y_i)$, where $i = 1, \dots, n - 1$, Algorithm 7.2 returns a fuzzy predictive distribution (\cite{Vovk2005}, eq.~(7.7)) that is defined in terms of quantities $C_1, \ldots, C_{n-1}$.  Comparative evaluation requires a crisp predictive distribution, for which we use the empirical distribution of $C_1, \ldots, C_{n-1}$, which adheres to the bounds imposed by the fuzzy distribution.\footnote{Here and in Section \ref{sec:computing}, we adopt the convention in Vovk~et al.~(\cite{Vovk2005}, Section 7.2) and assume that the size of the training set is $n - 1$, rather than $n$, to allow for direct references to material therein.  The respective crisp CDF is given by $F(y) = i/n$ for $y \in (C_{(i)}, C_{(i+1)})$ and $i = 0, 1, \ldots, n - 1$, and $F(y) = i''/n$ for $y = C_{(i)}$ and $i = 1, \ldots, n - 1$, where $C_{(0)} = - \infty$, $C_{(1)} \leq \cdots \leq C_{(n-1)}$ are the order statistics of $C_1, \ldots, C_{n-1}$, $C_{(n)} = \infty$ and $i'' = \max \{ \, j : C_{(j)} = C_{(i)} \}$.  For related discussion and alternative choices of a crisp CDF that is compatible with the fuzzy CDF, see Section 2 of Bostr{\"o}m et al.~\cite{Bostrom2021} and Section 5 of Vovk et al.~\cite{Vovk2020b}.}  For moderate to large training sets and $x$ the value of the model output at hand, $C_i$ typically is very close to $\hat{y} + y_i - \hat{y}_i$, where $\hat{y}$ and $\hat{y}_i$ are least squares point predictions based on $x$ and $x_i$, respectively (\cite{Vovk2005}, Section 7.3.4).

Finally, we consider the state-of-the-art approach to uncertainty quantification in weather prediction, namely, ensemble forecasts \cite{Gneiting2005a, Leutbecher2008, Palmer2000}, which are input-based methods.  Specifically, we use the world leading ECMWF Integrated Forecast System (IFS, \url{https://www.ecmwf.int/en/forecasts}), which comprises 51 NWP runs, namely, a control run and 50 perturbed members \cite{Molteni1996}.  The control run is based on the best estimate of the initial state of the atmosphere, and the perturbed members start from slightly different states that represent uncertainty. Even a single NWP model run, such as the HRES run, is computationally very expensive, and computing power is the limiting factor to improving model resolution.  Despite having coarser resolution, an ensemble typically requires 10 to 15 times more computing power than a single run \cite{Bauer2015}.  In contrast, the implementation of the output-based CP and EasyUQ methods is fast, with hardly any resources needed beyond a single NWP model run.

\renewcommand{\arraystretch}{1.1}
\begin{table}
\centering
\caption{Predictive performance in terms of mean $\CRPS$ for WeatherBench forecasts of upper air temperature at lead times of three and five days, in degrees Kelvin.  The evaluation period comprises calendar years 2017 and 2018.  CP and EasyUQ generate predictive CDFs that are fitted at each grid point individually, based on training data from 2010 through 2016.  Forecasts are issued twice daily, and scores are averaged over 32 $\times$ 64 grid points, for a total of 2,990,080 forecast cases.  \label{tab:CRPS_WB}}
\bigskip
\small
\begin{tabular}{llcc}
\toprule
\multicolumn{2}{c}{Forecast}                          & \multicolumn{2}{c}{$\CRPS$} \\
\multicolumn{1}{c}{Type} & \multicolumn{1}{c}{Method} & Three Days & Five Days \\
\midrule
Single-valued   & Climatology           & 2.904  & 2.904 \\
                & CNN                   & 2.365  & 2.782 \\
	            & HRES                  & 0.998  & 1.543 \\
\midrule     
Distributional  & CP on Climatology     & 2.055 & 2.055 \\
                & CP on CNN             & 1.673 & 1.955 \\
                & CP on HRES            & 0.731 & 1.123 \\  
\midrule
Distributional  & EasyUQ on Climatology & 2.038 & 2.038  \\
	            & EasyUQ on CNN         & 1.671 & 1.949  \\
	            & EasyUQ on HRES        & 0.736 & 1.122  \\
\midrule
Distributional  & ECMWF Ensemble        & 0.696 & 0.998  \\
\bottomrule
\end{tabular}
\end{table}

To compare CP and EasyUQ predictive CDFs to the respective single-valued forecasts we use the $\CRPS$ from \eqref{eq:CRPS} and recall that for single-valued forecasts the mean $\CRPS$ reduces to the mean absolute error (MAE).  As evaluation period, we take calendar years 2017 and 2018; for estimating the CP and EasyUQ predictive distributions, we use training data from calendar years 2010 through 2016 and proceed grid point by grid point.  The corresponding results are provided in Table \ref{tab:CRPS_WB}.  Not surprisingly, the ECMWF ensemble forecast has the lowest mean $\CRPS$.  However, CP and EasyUQ based on the HRES model output results in promising $\CRPS$ values, even though the methods require considerably less computing time and resources.

The CP and EasyUQ predictive distributions show nearly identical predictive performance.  To understand this behavior, we note that in the case of temperature, Gaussian predictive distributions with fixed variance typically are very adequate (see, e.g., \cite{Gneiting2005b}, Table 3).  In this light, key requirements of CP in the form of the LSPM (namely, fixed spread and fixed shape of the predictive distributions) and EasyUQ (namely, isotonicity) are reasonably met.  While EasyUQ generates predictive distribution that vary in spread and shape, the variations remain modest (Fig.~\ref{fig:nutshell}c--f), and the CP distributions, which essentially are translates of each other, are competitive.

The subsequent case study turns to a weather variable that is not covered by the WeatherBench challenge, but serves to illuminate and highlight difference between the CP and EasyUQ techniques.

\subsection{Illustration on precipitation forecasts}  \label{sec:BasicPrecip}  

Precipitation accumulation is generally considered the ``most difficult weather variable to forecast'' \cite{EbertUphoff2023}.  Indeed, the uncertainty quantification for deterministic forecasts of precipitation is more challenging than for temperature, since precipitation accumulation follows a mixture distribution with a point mass at zero --- for no precipitation --- and a continuous part on the positive real numbers.  Applying CP without corrections is bound to transfer mass to negative values of precipitation accumulation.  Taking advantage of knowledge about the outcome distribution, a natural remedy is to censor at zero and use the CDF
\begin{align*}  \label{eq:check_G}
G(y) = \begin{cases} 0, & y < 0, \\ F(y), & y \geq 0, \end{cases}
\end{align*}
in lieu of $F$.\footnote{In our experiments, we train without consideration of censoring, and censor at zero ex post.  For a nonnegative outcome, such a procedure guarantees improvement, in the technical sense that $\CRPS(G,y) \leq \CRPS(F,y)$ for all $y \geq 0$.  Alternatively, one might take censoring into account during training already.  However, methods of this latter type are more complex to implement, and improvements in CRPS cannot be guaranteed out-of-sample.}  In contrast, the EasyUQ predictive distributions reflect the nonnegativity of the outcomes in the training data, without any need for adaptation.

We now investigate the performance of CP and EasyUQ within the experimental setup from Henzi et al.~\cite{Henzi2021a}, taking forecasts and observations of 24-hour accumulated precipitation from 6 January 2007 through 1 January 2017 at Frankfurt airport, Germany.  Just as in the WeatherBench example, we consider a weekly climatology, the HRES forecast, and the 51 member NWP ensemble from ECMWF.  The weekly climatology is computed over the period 2007 to 2014, which is the same period that is used for CP and EasyUQ training.  The evaluation period comprises calendar years 2015 and 2016.  Table \ref{tab:CRPS_precip} shows the mean CRPS over the evaluation period for the various types of forecasts at lead times from one to five days.  Evidently, the climatological forecasts, along with their scores, do not depend on the lead time.  In contrast to the WeatherBench temperature example, EasyUQ outperforms CP for both Climatology and the HRES model output, and at all lead times.  While censoring improves the distributional forecasts from CP, the performance gap to EasyUQ remains pronounced.  EasyUQ on the HRES model output even outperforms the raw ECMWF ensemble at lead times of one and two days.\footnote{This is largely due to the fact that gridded ensemble predictions are compared against station observations.  To counter these effects, the ensemble forecast itself can be subjected to statistical postprocessing, i.e., the application of statistical methods to correct for biases and dispersion errors \cite{Gneiting2005b, Raftery2005}.  Parametric methods based on distributional regression \cite{Messner2014, Scheuerer2014} model the distribution of precipitation accumulation with censored logistic or censored generalized extreme value distributions.  An alternative approach is taken in Bayesian model averaging \cite{Sloughter2007}, which posits separate parametric forms for the probability of zero precipitation and the density at positive amounts.  Evidently, discrete-continuous mixture distributions considerably complicate model building and estimation, and great efforts are to be undertaken to find suitable parametric families for specific weather variables.  For a detailed performance comparison on the data on hand see Fig.~5 of Henzi et al.~\cite{Henzi2021a}, whose study also includes versions of IDR with multivariate covariates derived from the full ECMWF ensemble and suitable partial orders on them, an option alluded to at the end of Section \ref{sec:IDR}.  These yield improvements compared to both the raw ensemble forecast and EasyUQ on HRES, at the price of higher conceptual complexity, higher computational costs, and the need for access to the full ensemble, rather than single-valued HRES model output.}

\renewcommand{\arraystretch}{1.1}
\begin{table}
\centering
\caption{Predictive performance in terms of mean $\CRPS$ for forecasts of daily precipitation accumulation at Frankfurt airport at lead times from one to five days, in millimeter.  CP and EasyUQ generate predictive CDFs based on training data from 2007 through 2014.  The evaluation period comprises calendar years 2015 and 2016.  \label{tab:CRPS_precip}}
\bigskip
\small
\begin{tabular}{llccccc}
\toprule
\multicolumn{2}{c}{Forecast}                 & \multicolumn{5}{c}{$\CRPS$} \\
\multicolumn{1}{c}{Type} & \multicolumn{1}{c}{Method} 
                                             & 1 Day & 2 Days & 3 Days & 4 Days & 5 Days \\
\midrule
Single-valued   & Climatology                & 2.187 & 2.187 & 2.187 & 2.187 & 2.187 \\
                & HRES                       & 1.125 & 1.294 & 1.412 & 1.478 & 1.686 \\
\midrule  
Distributional  & CP on Climatology          & 1.382 & 1.382 & 1.382 & 1.382 & 1.382 \\
                & CP on HRES                 & 0.886 & 0.966 & 1.063 & 1.081 & 1.129 \\
                & Censored CP on Climatology & 1.324 & 1.324 & 1.324 & 1.324 & 1.324 \\
                & Censored CP on HRES        & 0.850 & 0.925 & 1.031 & 1.050 & 1.100 \\
\midrule
Distributional  & EasyUQ on Climatology      & 1.242 & 1.242 & 1.242 & 1.242 & 1.242 \\
	              & EasyUQ on HRES             & 0.732 & 0.803 & 0.876 & 0.945 & 1.001 \\
\midrule
Distributional  & ECMWF Ensemble             & 0.752 & 0.847 & 0.856 & 0.918 & 0.981 \\
\bottomrule
\end{tabular}
\end{table}

\begin{figure}[t]
\centering
\includegraphics[width=\textwidth]{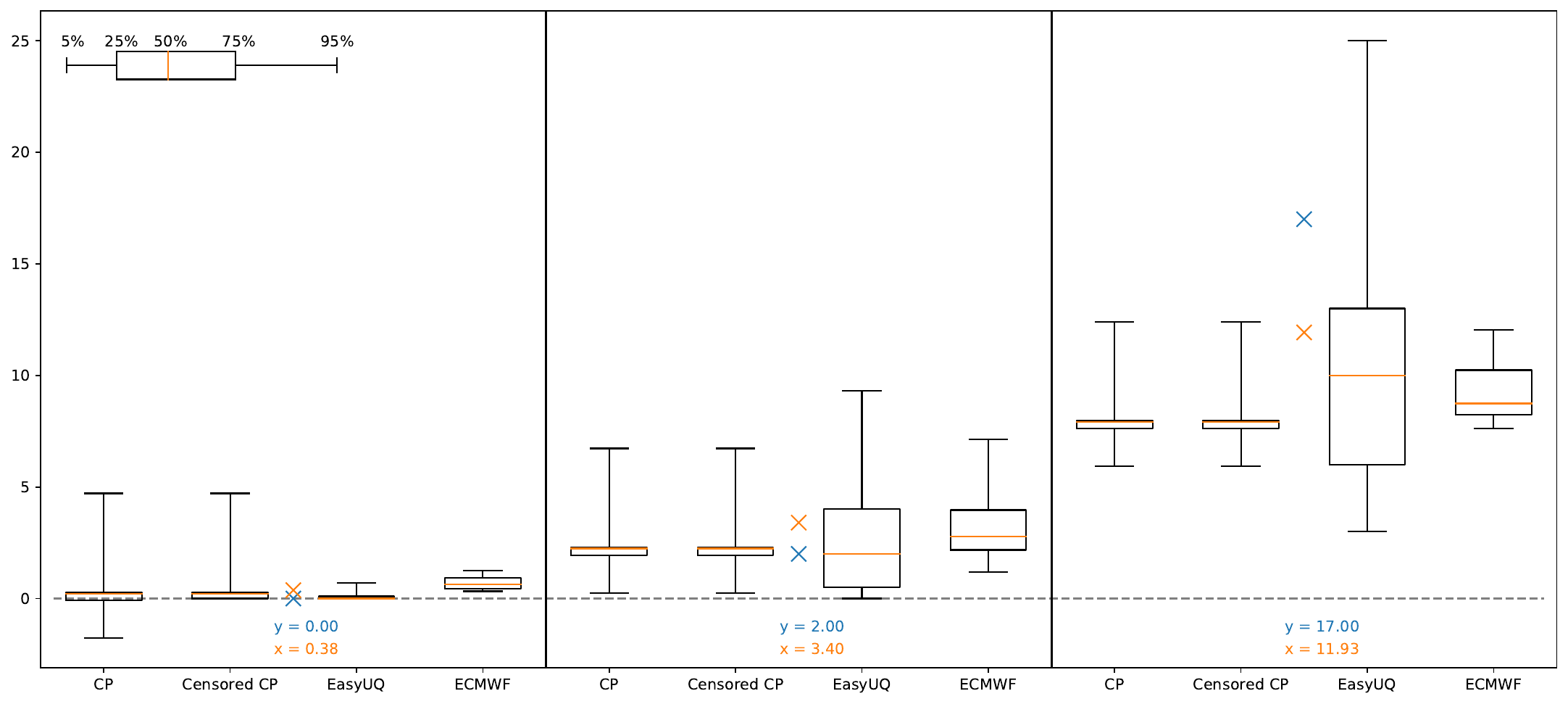}
\vspace{-5mm}
\caption{One-day ahead forecasts of daily precipitation accumulation at Frankfurt airport valid 23 January 2015 (left, HRES model output $x$ equal to 0.38, as indicated by the blue cross), 14 January 2015 (middle, $x$ = 3.40), and 21 February 2016 (right, $x$ = 11.93), in millimeter.  The predictive distributions for CP on HRES, Censored CP on HRES, EasyUQ on HRES, and ECMWF ensemble techniques are shown.  The observed precipitation accumulation was at $y$ = 0, $y$ = 2, and $y$ = 17 millimeter, respectively.  \label{fig:precip}}
\end{figure}

Figure \ref{fig:precip} provides a graphical comparison of CP on HRES, Censored CP on HRES, EasyUQ on HRES, and ECMWF ensemble forecasts at small ($x = 0.38$), moderate  ($x = 3.40$), and large ($x = 11.93$) values of the HRES model output $x$.  We see that the CP predictive distributions essentially are translates of each other, with mass potentially being transfered to negative values of precipitation accumulation, and censoring shifting any such mass to zero.  In contrast, the ECMWF ensemble and EasyUQ distributions do not have mass at negative values, and they vary in shape and scale.  However, while the ECMWF ensemble tends to show forecast distributions that are too narrow, as frequently observed in practice \cite{Gneiting2005a} and illustrated by the right-hand example, the EasyUQ distributions, which are based on the single-valued HRES forecast only, show what appears to be adequate spread.  Remarkably, and unlike any other method that we are aware of, EasyUQ achieves this desirable performance in its very basic form, without any need for implementation decisions, parameter tuning, or other forms of adaptation and intervention.

\section{Smooth EasyUQ}  \label{sec:Smooth}  

EasyUQ provides discrete predictive distributions with positive probability mass at the outcomes from the training archive.  For genuinely discrete outcomes, the variable of interest attains a small number of unique values only, and this is a desirable property.  For genuinely continuous variables, it is preferable to use continuous predictive distributions.  We now describe the Smooth EasyUQ technique, which turns the discrete basic EasyUQ CDFs into continuous Smooth EasyUQ CDFs with Lebesgue densities, while preserving isotonicity.  To achieve this, Smooth EasyUQ applies kernel smoothing, which requires implementation choices, unlike basic EasyUQ that does not require any tuning.  However, we provide default options.

\subsection{Smooth EasyUQ: Kernel smoothing under isotonicity preservation}  \label{sec:kernel}

Our goal is to transform the discrete basic EasyUQ CDFs $\hat{F}_x$ from \eqref{eq:hat_F} into smooth predictive CDFs $\check{F}_x$ that admit Lebesgue densities $\check{f}_x$, without abandoning the order relations honored by the basic technique.  To this end, we define the Smooth EasyUQ CDF as
\begin{equation}  \label{eq:check_FF}
\check{F}_x(y) = \int_{-\infty}^\infty \hat{F}_x(t) \, K_h(y-t) \, \textrm{d}t, 
\end{equation}
where $K_h(u) = (1/h) \, \kappa(u/h)$ for a smooth probability density function or kernel $\kappa$, such as a standardized Gaussian or Student-$t$ density, with bandwidth $h > 0$.  While the convolution approach in \eqref{eq:check_FF} is perfectly general for the smoothing of CDFs, we henceforth focus the presentation on EasyUQ.  The choice of the kernel and the bandwidth are critical, and we tend to their selection in the next section, where we introduce multiple one-fit grid search as a computationally efficient alternative to cross-validation.    

\begin{figure}[t]
\centering
\includegraphics[width=0.9\textwidth]{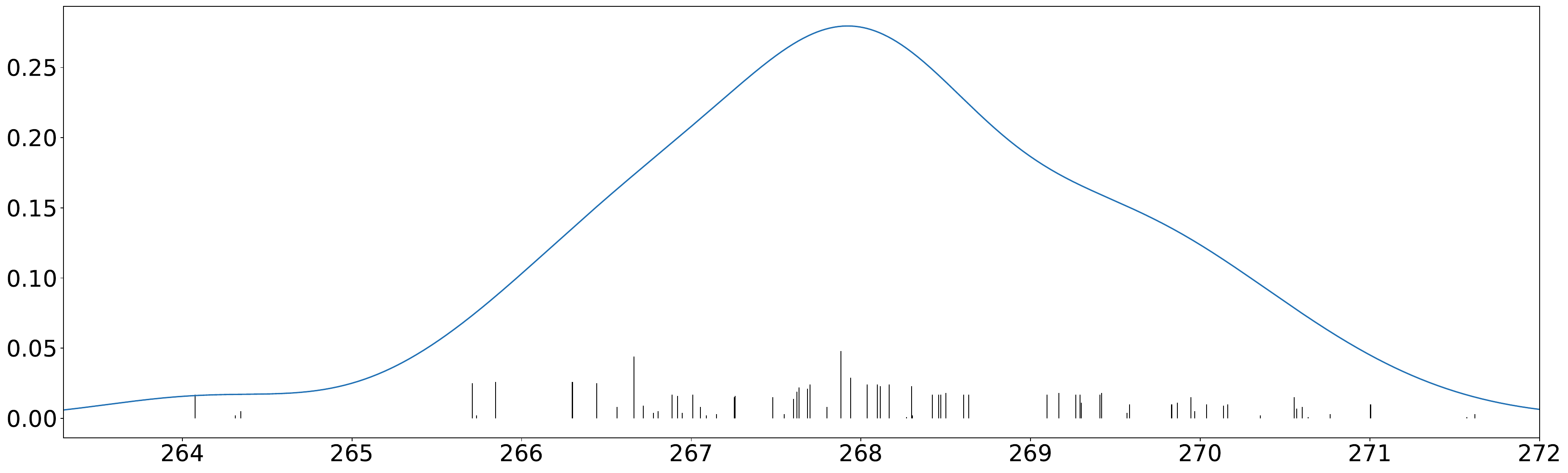}
\caption{Smooth EasyUQ predictive density \eqref{eq:check_f} in the WeatherBench example from Fig.~\ref{fig:nutshell}f) at HRES model output $x$ equal to 268 degrees Kelvin.  The vertical bars show the weights $w_1(x), \ldots, w_k(x)$ that the discrete EasyUQ distribution $\hat{F}_x$ assigns to the unique values $\tilde{y}_1 < \cdots < \tilde{y}_k$ of the outcomes in the training set, where $k$ = 76.  \label{fig:check_f}}
\end{figure}

For now, recall that $\hat{F}_{x}(y)$ from \eqref{eq:hat_F} is a step function with possible jumps at the unique values $\tilde{y}_1 < \cdots < \tilde{y}_k$ of the outcomes $y_1, \ldots, y_n$ in the training set.  Hence, we can write \eqref{eq:check_FF} as
\begin{align*}
\check{F}_x(y) = \sum_{j = 1}^k \hat{F}_x(\tilde{y}_j) \int_{\tilde{y}_{j}}^{\tilde{y}_{j+1}} K_h(y-t) \, \textrm{d}t, 
\end{align*}
where $\tilde{y}_{k + 1} = \infty$.  To compute the density $\check{f}_x = \check{F}'_x$, we set $\tilde{y}_0 = - \infty$, note that $\hat{F}_x$ assigns mass $w_j(x) = \hat{F}_x(\tilde{y}_j) - \hat{F}_x(\tilde{y}_{j-1})$ to $\tilde{y}_j$, and find that  
\begin{align}  \label{eq:check_f}
\check{f}_x(y) = \sum_{j = 1}^k \hat{F}_x(\tilde{y}_j) \, [K_h(y-\tilde{y}_j) - K_h(y-\tilde{y}_{j+1})] = \sum_{j = 1}^k \, w_j(x) \, K_h(y-\tilde{y}_j).
\end{align}
In words, the Smooth EasyUQ density $\check{f}_{x}$ from \eqref{eq:check_f} arises as a kernel smoothing of the discrete probability measure that corresponds to $\hat{F}_x$ and assigns weight $w_j(x)$ to $\tilde{y}_j$.  Consequently, $\check{f}_x$ is a probability density function, $\check{F}_x$ is a proper CDF, and, notably, Smooth EasyUQ preserves the stochastic ordering of the basic EasyUQ estimates.  In Fig.~\ref{fig:check_f} we illustrate the interpretation of the Smooth EasyUQ density as a kernel smoothing of the EasyUQ point masses $w_j(x)$ on the WeatherBench example.

Subject to mild regularity conditions, the asymptotic consistency of EasyUQ carries over to Smooth EasyUQ.  To demonstrate this, we prove a general consistency theorem for estimates of conditional CDFs in Appendix \ref{app:consistency}.  Here, we sketch how the result applies in the special case of Smooth EasyUQ.  Specifically, let $\check{F}_{x;n}$ denote the Smooth EasyUQ estimator from \eqref{eq:check_FF}, where the basic estimate $\hat{F}_x$ is trained on a sample of size $n$ from a population with true conditional CDFs $F_x$ that are H\"older continuous with constants $\alpha \in (0,1]$ and  $L > 0$.  Suppose that the function $K_h$ in \eqref{eq:check_FF} uses a kernel $\kappa$ with a finite absolute moment $c_{\kappa,\alpha}$ of order $\alpha$, and allow for the bandwidth $h_n > 0$ to vary with the sample size.  Crucially, we assume that the sup-error of the basic estimate $\hat{F}_x$ at sample size $n$ is upper bounded by $\varepsilon_n$ in the asymptotic sense specified by \eqref{eq:hat_F_consistent}; a choice of $\varepsilon_n \sim (\log(n)/n)^{1/3}$ applies to EasyUQ, if on an interval $(a,b)$ the covariate values $x_1, \dots, x_n$ are sufficiently dense and the conditional CDFs are Lipschitz continuous in $x$.  Then Thm.~\ref{thm:smoothing} implies that, for some sequence of $\delta_n > 0$ converging to zero,
\begin{align}  \label{eq:consistency} 
\lim_{n \to \infty} \myP \left( \: \sup_{y \in \real, \, x \in (a + \delta_n, b - \delta_n)} \left| \check{F}_{x;n}(y) - F_x(y) \right| \geq \varepsilon_n + L c_{\kappa,\alpha} h_n^\alpha \right) = 0.
\end{align}
If $h_n^\alpha = \cO(\varepsilon_n)$ the Smooth EasyUQ CDFs converge with the same rate as the basic EasyUQ CDFs.  For details, proofs, and discussion see Appendix \ref{app:consistency}.  In a nutshell, smoothness conditions on the true conditional CDFs are essential and unproblematic, as one should not be replacing the basic version of EasyUQ by Smooth EasyUQ in practice, unless there is subject matter indication of an absolutely continuous distribution of the outcome.  For instance, in the precipitation forecasting example from Section \ref{sec:BasicPrecip}, basic EasyUQ outperforms Smooth EasyUQ and a censored version of it at all lead times, cf.~Tables \ref{tab:CRPS_precip} and \ref{tab:precip_density}. 

\subsection{Choice of kernel and bandwidth: Multiple one-fit grid search}  \label{sec:choice}

In order to compute the Smooth EasyUQ density $\check{f}_x$ from \eqref{eq:check_f} one needs to choose a kernel $\kappa$ and a bandwidth $h > 0$, to yield a mixture of translates of the density $K_h(u) = (1/h) \, \kappa(u/h)$.  While there is a rich literature on bandwidth selection for kernel density estimation and kernel regression (see, e.g., \cite{Koehler2014, Silverman1986}), caution is needed when applying established approaches to Smooth EasyUQ, due to the fact that smoothing is applied to estimated conditional CDFs rather than raw data.

Furthermore, while the extant literature focuses on bandwidth selection for a fixed kernel, approaches of this type are restrictive for our purposes.  The Smooth EasyUQ density from \eqref{eq:check_f} inherits the tail behavior of the kernel $\kappa$, and so the properties of the kernel are of critical importance to the quality of the uncertainty quantification in the tails of the conditional distributions.  To allow for distinct tail behaviour, we use the Student-$t$ family and set $K_{\nu,h}(u) = (1/h) \, \kappa_\nu(u/h)$, where
\begin{align}  \label{eq:Student-t}
\kappa_\nu(y) = \frac{\Gamma((\nu+1)/2)} {(\pi\nu)^{1/2} \, \Gamma(\nu/2)} \left( 1 + \frac{y^2}{\nu} \right)^{-(\nu+1)/2}
\end{align}
is a standardized Student-$t$ probability density function with $\nu > 0$ degrees of freedom.  It is well known that the Student-$t$ distribution has a finite first moment if $\nu > 1$ and a finite variance if $\nu > 2$.  In the limit as $\nu \to \infty$, we find that $\kappa_\nu(y) \to \kappa_\infty(y)$ uniformly in $y$, where $\kappa_\infty(y) = (2\pi)^{-1/2} \exp(-y^2/2)$ is the standard Gaussian density function, so the ubiquitous Gaussian kernel emerges as a limit case in \eqref{eq:Student-t}. 

Turning to the choice of the tail parameter $\nu \in (0,\infty]$ and the bandwidth $h > 0$, we begin by discussing the latter.  A popular approach for bandwidth selection, both in kernel regression and kernel density estimation, is leave-one-out cross-validation.  Here the target criterion in terms of the bandwidth is
\begin{align}  \label{eq:CV}
\mathrm{CV}(h) = \frac{1}{n} \sum_{i = 1}^n \myS(\check{F}_{x_i,-i,h}, y_i),
\end{align}
where $\myS$ is a proper scoring rule, and $\check{F}_{x_i,-i,h}$ is the Smooth EasyUQ CDF with covariate $x_i$ and bandwidth $h$, estimated with all data from \eqref{eq:data} except for the $i$-th instance.  The optimization of the target criterion \eqref{eq:CV} either uses the CRPS as loss function $\myS$, as (implicitly) suggested for the estimation of conditional CDFs and quantile functions (see, e.g., \cite{Bowman1998}, p.~801 and \cite{Li2013}, p.~58) and yielding a target that is asymptotically equivalent to the integrated mean squared error (\cite{Henzi2021a}, Section S4), or the $\LogS$, as proposed for ensemble smoothing \cite{Brocker2008}.  We take the latter as default choice, since the $\LogS$ is much more sensitive to the choice of the bandwidth $h$ than the more robust $\CRPS$.

However, there are a number of caveats.  Empirical data typically are discrete to some extent, and might contain ties in the response variable, such as in the setting of Fig.~\ref{fig:check_f}, where there are only $m = 76$ unique values among the outcomes $y_1, \dots, y_n$, even though $\check{f}_x$ is estimated from a training archive of size $n = 5,114$.  In such cases, the optimal cross-validation bandwidth under the $\LogS$ may degenerate to $h = 0$, a problem that is also known in density estimation (\cite{Silverman1986}, pp.~51--55), in the estimation of Student-$t$ regression models \cite{Fernandez1999} and, in related form, in performance evaluation for forecast contests \cite{Kohonen2006, Quinonero2006}.  Another issue is that leave-one-out cross-validation is computationally expensive, as for each value of $h$ it requires the computation of $n$ distinct IDR solutions.  While a potential remedy is to remove a higher percentage of observations in each cross-validation step, we use a considerably faster approach, which we term one-fit grid search, that addresses both issues simultaneously.  

One-fit grid search avoids repeated fits of IDR and computes EasyUQ only once, namely, on the full sample from \eqref{eq:data}.  Specifically, given any fixed kernel $\kappa$, one-fit grid search finds the optimal bandwidth $h$ in terms of the target criterion
\begin{align}  \label{eq:OF}
\mathrm{OF}(h) = \frac{1}{n} \sum_{i = 1}^n \myS(\bar{F}_{x_i,-i,h}, y_i), 
\end{align}
where $\bar{F}_{x_i,-i,h}$ removes the unique value $\tilde{y}_j = y_i$ from the support of $\check{F}_{x_i}$ in \eqref{eq:check_FF}, by setting $w_j(x)$ in \eqref{eq:check_f} to zero and rescaling the remaining weights.  We choose the $\LogS$ as the default option for the loss function $\myS$ in the one-fit criterion \eqref{eq:OF}, and we use Brent's algorithm \cite{Brent1973} for optimization.  Effectively, one-fit grid search is a fast approximation to cross-validation, and when $n$ is small, leave-one-out cross-validation and the original criterion in \eqref{eq:CV} can be used instead, of course.  To choose a Student-$t$ kernel, we repeat the procedure, i.e., we consider values of $\nu \in \{ 2, 3, 4, 5, 10, 20, \infty \}$ in \eqref{eq:Student-t}, with $\nu = \infty$ yielding the Gaussian limit, apply one-fit grid search for each of these values, to find the respective optimal bandwidth $h$, and select the combination of $\nu$ and $h$ for which the target criterion \eqref{eq:OF} is smallest overall.  While being highly effective in our experience, multiple one-fit grid search is a crude approach, and we encourage further development. 

\subsection{Illustration on temperature and precipitation forecast examples}  \label{sec:SmoothWBPrecip}

For an initial illustration, we return to the WeatherBench challenge and the EasyUQ density in Figs.~\ref{fig:nutshell}f) and \ref{fig:check_f}, where $n = 5,114$ and $m = 76$, and multiple one-fit grid search with respect to the $\LogS$ yields parameter values $\nu = \infty$ and $h = 0.60$ in the kernel density \eqref{eq:Student-t}.  Considering the $32 \times 64 = 2,048$ grid points in WeatherBench and predictions three days ahead, the value of $\nu$ selected the most frequently for EasyUQ on the HRES model output, namely, $619$ times, is $\nu = 10$, with a median choice of $h = 0.49$.  For Smooth EasyUQ on Climatology and CNN $\nu = \infty$ was most frequently selected, namely $1,391$ and $1,361$ times with median choices of $h = 0.85$ and $h = 1.04$, respectively. 

A very simple, frequently used reference method for converting single-valued model output into a predictive density is the Single Gaussian technique \cite{Doubleday2020}.  It issues a Gaussian distribution with mean equal to the single-valued model output, and a constant variance that is optimal with respect to the mean $\LogS$ on a training set, which here we take to be the same as for EasyUQ.  Evidently, both Smooth EasyUQ and the Single Gaussian technique could be trained in terms of the $\CRPS$ as well.  {We also compare to the Smooth CP technique, which converts the discrete CP distributions to densities, as described in the next section. 

\renewcommand{\arraystretch}{1.1}
\begin{table}[t]
\centering
\caption{Predictive performance in terms of mean $\LogS$ and mean $\CRPS$ for WeatherBench density forecasts of upper air temperature at lead times of three and five days, in degrees Kelvin.  The evaluation period comprises calendar years 2017 and 2018.  The Single Gaussian, Smooth CP, and Smooth EasyUQ methods are trained at each grid point individually, based on data from 2010 through 2016.  Forecasts are issued twice daily, and scores are averaged over $32 \times 64$ grid points, for a total of 2,990,080 forecast cases.  \label{tab:LogS_WB}}
\bigskip
\small
\begin{tabular}{lcccc}
\toprule
Density Forecast  & \multicolumn{2}{c}{$\LogS$} & \multicolumn{2}{c}{$\CRPS$} \\
\midrule
\hfill Days Ahead              & Three & Five  & Three & Five \\
\midrule
Single Gaussian on Climatology & 2.578 & 2.578 & 2.060 & 2.060 \\
Single Gaussian on CNN         & 2.413 & 2.553 & 1.696 & 1.983 \\
Single Gaussian on HRES        & 1.694 & 2.073 & 0.748 & 1.153 \\
\midrule 
Smooth CP on Climatology       & 2.562 & 2.562 & 2.059 & 2.059 \\
Smooth CP on CNN               & 2.384 & 2.519 & 1.672 & 1.952 \\
Smooth CP on HRES              & 1.627 & 2.007 & 0.732 & 1.123 \\
\midrule
Smooth EasyUQ on Climatology   & 2.540 & 2.540 & 2.043 & 2.043 \\
Smooth EasyUQ on CNN           & 2.375 & 2.509 & 1.667 & 1.945 \\
Smooth EasyUQ on HRES          & 1.640 & 2.006 & 0.736 & 1.122 \\
\midrule
Smoothed ECMWF Ensemble        & 1.503 & 1.824 & 0.685 & 0.990 \\
\bottomrule
\end{tabular}
\end{table}

In Table \ref{tab:LogS_WB}, we evaluate Smooth EasyUQ, Smooth CP, and Single Gaussian density temperature forecasts in the WeatherBench setting.  For evaluation, we use both the $\CRPS$ and the $\LogS$.  Throughout, Smooth EasyUQ and Smooth CP outperform the Single Gaussian method, though they do not match the performance of the smoothed ECMWF ensemble forecast, which we construct as follows.  Let $\tilde{z}_1 < \cdots < \tilde{z}_k$ be the unique values of the ensemble members $z_1, \dots, z_l$ of an ensemble forecast of size $l$.  The smoothed ensemble CDF then is of the form \eqref{eq:check_FF} with mass $w_j = \frac{1}{l} \sum_{i = 1}^l \one (z_i = \tilde{z}_j)$ for $j = 1, \ldots, k$.  Interestingly, this is the same as Br\"{o}cker--Smith smoothing of ensemble forecasts (\cite{Brocker2008}, relations (19)--(21)) with parameters $a = 1$ and $r_1 = r_2 = s_2 = 0$ being fixed.  However, while Br\"{o}cker and Smith use a Gaussian kernel and optimize the bandwidth parameter only, we take a more flexible approach and consider values of $\nu \in \{ 2, 3, 4, 5, 10, 20, \infty \}$ for a Student-$t$ kernel, to find the optimal $\nu$ and bandwidth $h$ in terms of the $\LogS$.  Across the $2,048$ grid points, the most frequent choice is $\nu = 5$, namely, $743$ times, with a median bandwidth value of $h = 0.50$.

\renewcommand{\arraystretch}{1.1}
\begin{table}
\centering
\caption{Predictive performance in terms of mean $\CRPS$ for density forecasts of daily precipitation accumulation at Frankfurt airport at lead times from one to five days, in millimeter.  CP and EasyUQ generate predictive CDFs based on training data from 2007 through 2014.  The evaluation period comprises calendar years 2015 and 2016.  \label{tab:precip_density}}
\bigskip
\small
\begin{tabular}{lccccc}
\toprule
Density Forecast                 & 1 Day & 2 Days & 3 Days & 4 Days & 5 Days \\
\midrule  
Single Gaussian on HRES          & 1.244 & 1.380 & 1.547 & 1.577 & 1.724 \\
Censored Single Gaussian on HRES & 1.013 & 1.145 & 1.266 & 1.276 & 1.401 \\
\midrule
Smooth CP on HRES                & 0.886 & 0.971 & 1.064 & 1.087 & 1.132 \\
Censored Smooth CP on HRES       & 0.849 & 0.928 & 1.028 & 1.052 & 1.098 \\
\midrule
Smooth EasyUQ on HRES            & 0.760 & 0.828 & 0.901 & 0.968 & 1.033 \\
Censored Smooth EasyUQ on HRES   & 0.745 & 0.817 & 0.893 & 0.960 & 1.016 \\
\midrule
Smoothed ECMWF Ensemble          & 0.762 & 0.855 & 0.863 & 0.924 & 0.986 \\
Censored Smoothed ECMWF Ensemble & 0.750 & 0.850 & 0.860 & 0.921 & 0.984 \\
\bottomrule
\end{tabular}
\end{table}

While smoothing is warranted for temperature forecasts, it is problematic for forecasts of precipitation accumulation, due to the nonnegativity of the outcome and the point mass at zero.  Indeed, due to the kernel smoothing, the Smooth EasyUQ and smoothed ECMWF ensemble densities have mass on the negative halfaxis, unlike the discrete (basic) EasyUQ and (raw) ECMWF distributions, which are concentrated on the nonnegative halfaxis.  Nonetheless, Table \ref{tab:precip_density} compares the predictive performance of Single Gaussian, Smooth CP, Smooth EasyUQ, and smoothed ECMWF ensemble forecasts in the setting of Section \ref{sec:BasicPrecip}, in both original and censored variants. The results mirror the findings in Table \ref{tab:CRPS_precip}, in that censoring yields improvement and EasyUQ outperforms CP, whereas CP outperforms the Single Gaussian technique.

\subsection{Computational considerations}  \label{sec:computing}

We add a brief discussion of the computational complexity of output-based methods for uncertainty quantification.  For this comparison, we utilize the setting of Algorithm 7.2 in Vovk et al.~\cite{Vovk2005}, which requires predictive distributions for $m$ new values of $x$, based on a training set of size $n - 1$ with instances $(x_1, y_1), \ldots, (x_{n-1}, y_{n-1})$.  We report upper estimates of the computational complexity for the Single Gaussian technique, CP, and EasyUQ, considering both training (i.e., initial operations on the training data only) and inference (i.e., operations to be repeated for each new value).  For the simplistic Single Gaussian technique, training requires $\cO(n)$ operations and inference is straightforward.

For EasyUQ, the main effort lies in training, where the complexity is upper bounded by $\cO(n^2)$ operations \cite{Henzi2022}.  Training the EasyUQ CDFs only on a fixed grid of ordinates guarantees a cost reduction to $\cO(n \log n)$ operations, and Henzi et al.~\cite{Henzi2021a} describe approaches based on subset aggregation that reduce the computational burden for estimation. That said, the numerical experiments in our paper use the standard implementation throughout, without exception. For inference, each new value of $x$ requires the determination of its position within the unique values across $x_1, \ldots, x_{n-1}$, followed by interpolation of the trained EasyUQ CDFs at the predecessor and successor values, at up to $\cO(mn)$ operations.

For CP in the form of the studentized LSPM (\cite{Vovk2005}, Algorithm 7.2) essentially no training is required, but inference incurs $\cO(mn^2)$ operations. Residual-based approximations to CP, which are instances of split conformal predictive systems (\cite{Vovk2005}, Section 7.3.4, \cite{Vovk2018}), are much faster, shift the bulk of the cost to training at $\cO(n)$ operations, and yield nearly identical predictive performance to CP in our experience, except when training sets are small.

For both, CP and EasyUQ, we have implemented smoothing in ways that avoid cross-validation and honor the aforementioned bounds. Smooth EasyUQ uses one-fit grid search as developed in this paper. To generate the Smooth CP densities, we use kernel smoothing with a Gaussian kernel and bandwidth chosen according to Silverman's rule of thumb \cite{Silverman1986}, applied to the quantities $C_1, \ldots, C_{n-1}$ that arise for each new instance separately.

In the aforementioned experiments, we generally found the computational cost of EasyUQ to be nested in between the costs of CP and residual-based approximations to CP.\footnote{To provide intuition about computation times, we report mean run times for the Single Gaussian technique, CP, and EasyUQ applied to the HRES forecast in the setting of Table \ref{tab:CRPS_precip}, where the training set is of size 2,896 and the evaluation set of size 721.  The mean run time averaged over the five lead times is 0.005 seconds for the Single Gaussian technique, 0.45 seconds for CP, and 0.085 seconds for EasyUQ. We note that the computing time for CP on CPU is 33.64 seconds, but can be reduced to 0.45 seconds on GPU.  Evidently, the comparison faces the usual challenges, given that execution times depend on factors including but not limited to hardware architecture, disk speed, memory availability, and programming language and compiler used.  Specific to the situation at hand, we use code in Python, R, and C++, run some functions on GPU and others on CPU, and it is unlikely that every single of our implementations, which typically are based on packages, has been coded in the most efficient way.}  Compared to the enormous effort of running the HRES model, or even the input-based ECMWF ensemble method, which require the operational use of supercomputers, run times and computational costs for the output-based Single Gaussian, CP, and EasyUQ techniques are negligible.

\section{EasyUQ and neural networks}  \label{sec:NN} 

Neural networks and deep learning techniques have enabled unprecedented progress in predictive science.  However, as they ``can struggle to produce accurate uncertainties estimates~[\ldots]~there is active research directed toward this end'' (\cite{Baker2022}, p.~67)}, which has intensified in recent years \cite{Abdar2021, Chung2021, Duan2020, Gal2016, Immer2021, Kuleshov2018, Lakshminarayanan2017, Marx2022, Vovk2020a}.  We now discuss how EasyUQ and Smooth EasyUQ can be used to yield accurate uncertainty statements from neural networks.  Evidently, our methods apply in the ways described thus far, where single-valued model output is treated as given and fixed, with subsequent uncertainty quantification via EasyUQ or Smooth EasyUQ being a completely separate add-on, as illustrated on our temperature and precipitation examples.  In the context of neural networks, this means that the network parameters are optimized to yield single-valued output, and only then EasyUQ gets applied.  We now describe a more elaborate approach where we integrate our methods within the typical workflow of neural network training and evaluation. 

\subsection{Integrating EasyUQ into the workflow of neural network learning and hyperparameter optimization}  \label{sec:NN_workflow}

Neural networks and associated methods for uncertainty quantification are developed and evaluated in well-designed workflows that involve multiple splits of the available data into training, validation, and test sets.  For each split, the training set is used to learn basic neural network parameters, the validation set is used to tune hyperparameters, and the test set is used for out-of-sample evaluation.  Scores are then averaged over the tests sets across the splits, and methods with low mean score are preferred.  

Algorithm \ref{alg:NN} describes how Smooth EasyUQ can be implemented within this typical workflow of neural network learning and hyperparameter tuning.  In a nutshell, we treat the kernel parameters for Smooth EasyUQ, namely, the Student-$t$ parameter $\nu$ and the bandwidth $h$ as supplemental hyperparameters, and optimize over both the neural network hyperparameters and kernel parameters.  As the evaluation occurs out-of-sample, the issues associated with the choice of the kernel parameters discussed in Section \ref{sec:choice} are mitigated, unless a dataset is genuinely discrete, in which case even out-of-sample estimates of the bandwidth $h$ can degenerate to zero, thereby indicating that smoothing is ill-advised.  To handle even such ill-advised cases, we use a procedure that we call moderated grid search \cite{Walz2022}.  Specifically, we first check if using $\nu = 2$ or a Gaussian kernel results in a degeneration of the optimal bandwidth $h$ to zero, and if so, we use the latter with bandwidth chosen according to Silverman's rule of thumb \cite{Silverman1986}.  Otherwise, we consider values of $\nu \in \{ 2, 3, 4, 5, 10, 20, \infty \}$ in \eqref{eq:Student-t}, with $\nu = \infty$ yielding the Gaussian limit.  For each value of $\nu$, we use Brent's method \cite{Brent1973} to optimize the log score with respect to the bandwidth $h$ on the validation set, and choose the optimal combination of $\nu$ and $h$.  Once network hyperparameters and kernel parameters have been determined, we re-learn the neural network on the combined training and validation sets, using the optimized hyperparameters, and apply EasyUQ on the re-learned single-valued neural network output.  Finally, we apply Smooth EasyUQ based on the re-learned EasyUQ solution and the selected kernel parameters, to yield density forecasts on the test set.

\begin{algorithm}[t]
\caption{Integration of Smooth EasyUQ into the workflow of neural network training and hyperparameter tuning.  The procedure returns the mean score of the Smooth EasyUQ predictions across data splits. \label{alg:NN}}
\begin{algorithmic}[1]
	\For {split in mysplit}
	\State separate data into training set, validation set, and test set 	
	\For {hyperpar in myhyperpar}
	\State learn neural network with hyperpar on training set
	\State use neural network output to fit basic EasyUQ on training set
	\State use moderated grid search to select EasyUQ parameters $\nu$ and $h$ 
	\State save selected $(\nu, h)$ and mean score on validation set 
	\EndFor
	\State select best hyperpar and associated $(\nu,h)$, based on smallest mean score
	\State re-learn network with best hyperpar on combined training and validation sets
	\State use re-learned neural network output to re-fit basic EasyUQ on combined training and validation sets
	\State use Smooth EasyUQ based on re-fitted EasyUQ with best $(\nu,h)$ for predictions on test set
	\State save scores on test set 
	\EndFor 
	\State return mean score across splits 
\end{algorithmic}
\end{algorithm}	

While optimization could be performed with respect to the $\CRPS$, the $\LogS$, or any other suitable proper scoring rule, we follow the machine learning literature, where benchmarking is typically in terms of the $\LogS$.  The $\CRPS$ serves as an attractive alternative, much in line with recent developments in neural network training, where optimization is performed with respect to the $\CRPS$ \cite{D'Isanto2018, Rasp2017}.  Its use becomes essential in simplified versions of Algorithm \ref{alg:NN} that work with \add{the discrete} basic EasyUQ \add{distributions} rather than Smooth EasyUQ \add{densities}. 

\subsection{Application in benchmark settings from machine learning}  \label{sec:benchmark}

As no\-ted, our intent is to compare Smooth EasyUQ in the integrated version of Algorithm \ref{alg:NN} to extant, state of the art methods for uncertainty quantification from the statistical and machine learning literatures.   The comparison is on ten datasets for regression tasks using the experimental setup proposed and developed by Hernandez-Lobato and Adams \cite{HernandezLobato2015}, Gal and Gharahmani \cite{Gal2016}, Lakshminarayanan et al.~\cite{Lakshminarayanan2017}, and Duan et al.~\cite{Duan2020}.  Characteristics of the ten datasets are summarized in Table \ref{tab:LogS_ML}, including the size of the datasets, the number of unique outcomes, and the dimension of the input space for the regression problem. 

\begin{sidewaystable}
\centering
\caption{Characteristics of datasets and predictive performance for competing methods of uncertainty quantification in regression problems, in terms of the mean logarithmic score ($\LogS$) in a popular benchmark setting from machine learning \cite{Duan2020, Gal2016, HernandezLobato2015, Lakshminarayanan2017}.  For each dataset, we show size, number of unique outcomes, and dimension of the input (covariate or feature) space.  Italics indicate discrete datasets where the number of unique outcomes is small.  For each method, we report the mean $\LogS$ from the reference stated, with further details provided in Section \ref{sec:benchmark}.  For each of the lower three blocks of comparable methods, the best (lowest) mean score is set in {\it blue} color.  Two scores are numerically infinite; missing scores are marked NA.  \label{tab:LogS_ML}}
\bigskip
\scalebox{0.97}{
\begin{tabular}{lrrrrrrrrrr}
\toprule
Method / Dataset                                     & Boston   & Concrete &   Energy &     Kin8nm  &   \em Naval &    Power &  Protein &  \em Wine &    Yacht &   \em Year \\
\toprule
Size                                                 &      506 &    1,030 &      768 &       8,192 &  \em 11,934 &    9,568 &   45,730 & \em 1,599 &      308 & \em 515,345 \\ 
Unique Outcomes                                      &      229 &      845 &      586 &       8,191 &      \em 51 &    4,836 &   15,903 &     \em 6 &      258 &      \em 89 \\
Dimension Input Space                                &       13 &        8 &        8 &           8 &      \em 16 &        4 &        9 &    \em 11 &        6 &      \em 90 \\    
\toprule 
Distributional Forest \hfill \cite{Duan2020}         &     2.67 &     3.38 &     1.53 &     $-0.40$ &     $-4.84$ &     2.68 &     2.59 &      1.05 &     2.94 &          NA \\
GAMLSS \hfill \cite{Duan2020}                        &     2.73 &     3.24 &     1.24 &     $-0.26$ &     $-5.56$ &     2.86 &     3.00 &      0.97 &     0.80 &          NA \\
GP Regression \hfill \cite{Duan2020}                 &     2.37 &     3.03 &     0.66 &     $-1.11$ &     $-4.98$ &     2.81 &     2.89 &      0.95 &     0.10 &          NA \\ 
NGBoost \hfill \cite{Duan2020}                       &     2.43 &     3.04 &     0.60 &     $-0.49$ &     $-5.34$ &     2.79 &     2.81 &      0.91 &     0.20 &        3.43 \\ 
\midrule
40 Deep Ensembles \hfill \cite{Lakshminarayanan2017} & \it 2.41 &     3.06 &     1.38 & \it $-$1.20 &     $-5.36$ &     2.79 &     2.83 &      0.94 & \it 1.18 &        3.35 \\
40 Laplace \hfill \cite{Walz2022}                    &     2.65 &     3.14 &     1.27 &     $-1.00$ &          NA &     2.87 &     2.90 &      0.97 &     1.97 &        3.61 \\
40 Single Gaussian \hfill \cite{Walz2022}            &     2.78 &     3.20 &     1.14 &     $-1.03$ &     $-5.37$ &     2.83 &     2.93 &      0.98 &     2.11 &        3.61 \\ 
40 Smooth CP \hfill \cite{Walz2022}                  &     2.89 &     3.14 &     1.20 &     $-1.00$ &     $-5.52$ &     2.85 &     2.88 &      0.97 &     1.88 &          NA \\
40 Smooth EasyUQ \hfill \cite{Walz2022}              &     2.83 & \it 3.04 & \it 0.79 &     $-1.05$ & \it $-6.51$ & \it 2.77 & \it 2.48 &  \it 0.48 &     1.36 &    \it 3.24 \\
\midrule
400 MC Dropout \hfill \cite{Gal2016}                 & \it 2.46 &     3.04 &     1.99 &     $-0.95$ &     $-3.80$ &     2.80 &     2.89 &  \it 0.93 &     1.55 &        3.59 \\ 
400 Laplace \hfill \cite{Walz2022}                   &     2.61 &     3.07 &     0.80 &     $-1.11$ &          NA &     2.83 &     2.87 &      1.04 &     1.18 &        3.61 \\
400 Single Gaussian \hfill \cite{Walz2022}           &     3.41 &     3.32 &     0.85 &     $-1.09$ &     $-6.32$ &     2.81 &     2.87 &      1.38 &     2.04 &        3.61 \\ 
400 Smooth CP \hfill \cite{Walz2022}                 &     2.87 &     3.05 &     0.83 &     $-1.09$ &     $-6.65$ &     2.78 &     2.84 &      1.01 &     1.03 &          NA \\
400 Smooth EasyUQ \hfill \cite{Walz2022}             & \it 2.46 & \it 2.94 & \it 0.55 & \it $-1.13$ & \it $-7.51$ & \it 2.75 & \it 2.41 &      1.07 & \it 0.85 &    \it 3.24 \\
\midrule 
2L MC Dropout \hfill \cite{Gal2016}                  & \it 2.34 &     2.82 &     1.48 &     $-1.10$ &     $-4.32$ & \it 2.67 &     2.70 &  \it 0.90 &     1.37 &          NA \\
2L Laplace \hfill \cite{Walz2022}                    &     2.57 &     2.98 &     0.56 &     $-1.13$ &          NA &     2.76 &     2.81 &      1.22 &     1.24 &        3.60 \\
2L Single Gaussian \hfill \cite{Walz2022}            & $\infty$ &     3.78 &     0.74 &     $-0.96$ &     $-7.19$ &     2.76 &     2.77 &     10.51 & $\infty$ &        3.61 \\ 
2L Smooth CP \hfill \cite{Walz2022}                  &     2.66 &     2.94 &     0.63 &     $-1.18$ &     $-7.33$ &     2.70 &     2.67 &      1.01 &     0.74 &          NA \\
2L Smooth EasyUQ \hfill \cite{Walz2022}              &     2.49 & \it 2.71 & \it 0.36 & \it $-$1.21 & \it $-$8.20 & \it 2.67 & \it 2.30 &      0.95 & \it 0.50 &    \it 3.23 \\
\bottomrule
\end{tabular}
}
\end{sidewaystable}

Each dataset is randomly split 20 times into training (72\%), validation (18\%), and test (10\%) sets.  However, for the larger datasets, Protein and Year, the train-test split is repeated only five and a single time(s), respectively.  After finding the optimal set of (hyper)parameters, methods are re-trained on the combined training and validation set (90\%) and the resulting predictions are evaluated on the held-out test set (10\%).  We use the exact same splits as the extant literature in the implementation from \url{https://github.com/yaringal/DropoutUncertaintyExps}, and the final score is obtained by computing the average score over the splits.

Following the literature, we consider four techniques for the direct generation of conditional predictive distributions that do not use neural networks, namely, a semiparametric variant of the distributional forest technique \cite{Duan2020, Schlosser2019}, generalized additive models for location, scale and shape (GAMLSS, \cite{Stasinopoulos2007}), Gaussian process (GP) regression \cite{Rasmussen2005}, and natural gradient boosting (NGBoost, \cite{Duan2020}).  We adopt the exact implementation choices of \cite{Duan2020} for these techniques, which in some cases involve smoothing.  Except for NGBoost, scores for the Year dataset are unavailable (NA), in part, because methods fail to be computationally feasible for a dataset of this size.  

The remaining methods considered in Table \ref{tab:LogS_ML} are based on neural networks, and we adopt the network architectures proposed by Hernandez-Lobato and Adams \cite{HernandezLobato2015} and Gal and Ghahramani \cite{Gal2016}.  Specifically, we use the ReLU nonlinearity and either a single or two hidden layers, containing 50 hidden units for the smaller datasets, and 100 hidden units for the larger Protein and Year datasets.  To tune the network hyperparameters, namely, the regularization parameter $\lambda$ and the batch size, we use grid search.  Thus, the nested hyperparameter selection in the Smooth EasyUQ Algorithm \ref{alg:NN} finds a best combination of $\lambda$, the batch size, $\nu$, and $h$ by optimizing the mean $\LogS$.  Our intent is to compare EasyUQ and Smooth EasyUQ to state of the art methods for uncertainty quantification from machine learning, namely, Monte Carlo (MC) Dropout \cite{Gal2016} and Deep Ensembles \cite{Lakshminarayanan2017}, which perform uncertainty quantification directly within the workflow of neural network fitting.  Furthermore, these methods are input-based, i.e., they require access to, and operate on, the original covariate or feature vector.  As seen in the table, the dimensionality of the input space in the benchmark problems varies between 4 and 90.

In contrast, EasyUQ, CP, and the Single Gausian technique operate on the basis of the final model output only, and so can be applied without the original, potentially high-dimensional covariate or feature vector being available.  For CP we adapt our previously described implementation with further refined splits into training (57.6\%), calibration (14.4\%), validation (18\%), and test (10\%) sets.  Smooth CP uses the respective variant of Algorithm \ref{alg:NN}.  An intermediary role between input-based and output-based methods is assumed by the recently developed Laplace approach \cite{Immer2021, Ritter2018}, which leverages scalable Laplace approximations based on weights of the trained network.  For our numerical experiments we use the \texttt{laplace} software library for PyTorch \cite{Daxberger2021}.

A critical implementation decision in the intended comparisons is the number of training epochs in learning the neural network.\footnote{\add{For the purposes of this comparison, the number of training epochs needs to be fixed.  In practice, the number of epochs could be treated as a further hyperparameter and determined on the validation set.}}  While the original setup specifies 40 training epochs \cite{HernandezLobato2015}, MC Dropout uses 400 or, in the 2-layer configuration, 4,000 iterations \cite{Gal2016}.  Therefore, to enable proper comparison, we apply the competing methods in three distinct neural network configurations, namely, a single-layer network with 40 training epochs (prefix 40 in Tables \ref{tab:LogS_ML} and \ref{tab:CRPS_ML}), a single-layer network with 400 training epochs (prefix 400), and a 2-layer architecture with 4,000 training epochs (prefix 2L).  In Tables \ref{tab:LogS_ML} and \ref{tab:CRPS_ML}, key comparisons between techniques for uncertainty quantification then are within the respective three groups of methods, for which the neural network configurations used are identical.  

\subsection{Comparison of predictive performance}  \label{sec:results}  

We assess the predictive performance of EasyUQ and Smooth EasyUQ and other methods for probabilistic forecasting and uncertainty quantification, by comparing the mean $\LogS$ in Table \ref{tab:LogS_ML}.  We use the $\LogS$ from \eqref{eq:LogS} in negative orientation, so smaller values correspond to better performance.  Evidently, the use of the $\LogS$, which is customary in machine learning, prevents comparisons to the basic versions of EasyUQ and CP, to which we turn in Table \ref{tab:CRPS_ML}.  

A first insight from Table \ref{tab:LogS_ML} is that, generally, the methods in the second, third, and fourth block, which are based on neural networks, perform better relative to the direct, non neural network based methods in the first block (from top to bottom).  Thus, we focus attention on the comparison of distinct methods for uncertainty quantification in neural networks, namely, Deep Ensembles \cite{Lakshminarayanan2017} or MC droput \cite{Gal2016}, the Laplace approach \cite{Ritter2018}, the Single Gaussian technique, Smooth CP, and Smooth EasyUQ.  The 2-layer architecture generally improves results, compared to using a single layer for the neural network.  Smooth EasyUQ dominates the Single Gaussian and Smooth CP techniques and generally yields lower mean $\LogS$ than Deep Ensembles, MC Dropout, or the Laplace approach.  In 24 of the $3 \times 10 = 30$ five-fold comparisons across the bottom three blocks, Smooth EasyUQ achieves or shares the top score.  For eight of the ten datasets considered, the best performance across all 19 methods considered, including both neural network and non-neural network based techniques, is achieved or shared by Smooth EasyUQ under the 2-layer network architecture.  While this is not an exhaustive evaluation and no single method dominates universally, we note that Smooth EasyUQ is highly competitive with state of the art techniques for uncertainty quantification from machine learning. 

\begin{sidewaystable}
\centering
\caption{Predictive performance for competing methods of uncertainty quantification in regression problems, in terms of the mean $\CRPS$ in a popular benchmark setting from machine learning \cite{Duan2020, Gal2016, HernandezLobato2015, Lakshminarayanan2017}.  For each dataset, we show size, number of unique outcomes, and dimension of the input (covariate or feature) space.  Italics indicate discrete datasets where the number of unique outcomes is small.  For Kin8mn and Naval the mean $\CRPS$ has been multiplied by factors of 10 and 1,000, respectively.  For each block of comparable methods, the best (lowest) mean score is set in {\it blue} color.  For details see Section \ref{sec:benchmark}.  \label{tab:CRPS_ML}}
\bigskip
\scalebox{0.97}{
\begin{tabular}{lrrrrrrrrrr}
\toprule
Method / Dataset       &   Boston & Concrete &   Energy &   Kin8nm &  \em Naval &    Power &  Protein &  \em Wine &    Yacht &    \em Year \\
\toprule
Size                   &      506 &    1,030 &      768 &    8,192 & \em 11,934 &    9,568 &   45,730 & \em 1,599 &      308 & \em 515,345 \\ 
Unique Outcomes        &      229 &      845 &      586 &    8,191 &     \em 51 &    4,836 &   15,903 &     \em 6 &      258 &      \em 89 \\
Dimension Input Space  &       13 &        8 &        8 &        8 &     \em 16 &        4 &        9 &    \em 11 &        6 &      \em 90 \\    
\toprule 
40 Deep Ensembles      & \it 1.59 &     3.04 &     0.78 & \it 0.48 &   \it 0.41 &     2.23 &     2.40 &      0.34 & \it 0.45 &        4.35 \\ 
40 Laplace             &     1.71 &     3.02 &     0.45 &     0.49 &         NA &     2.46 &     2.46 &      0.35 &     0.80 &        4.72 \\ 
40 Single Gaussian     &     1.72 &     3.03 &     0.41 & \it 0.48 &       0.66 &     2.24 &     2.48 &      0.35 &     0.83 &        4.72 \\
40 CP                  &     1.73 &     3.04 &     0.45 &     0.49 &       0.58 &     2.24 &     2.47 &      0.36 &     0.90 &          NA \\
40 Smooth CP           &     1.74 &     3.05 &     0.45 &     0.49 &       0.58 &     2.24 &     2.47 &      0.36 &     0.91 &          NA \\
40 EasyUQ              &     1.69 &     2.94 &     0.34 & \it 0.48 &       0.54 &     2.21 &     2.22 &  \it 0.31 &     0.66 &        4.35 \\
40 Smooth EasyUQ       &     1.64 & \it 2.89 & \it 0.33 & \it 0.48 &       0.55 & \it 2.20 & \it 2.20 &      0.32 &     0.64 &    \it 4.34 \\
\midrule 
400 MC Dropout         & \it 1.56 &     2.79 &     0.37 &     0.48 &       1.22 &     2.21 &     2.40 &  \it 0.35 &     0.57 &        4.73 \\
400 Laplace            &     1.66 &     2.67 &     0.29 & \it 0.44 &         NA &     2.17 &     2.36 &      0.37 &     0.41 &        4.73 \\
400 Single Gaussian    &     1.61 &     2.72 &     0.29 & \it 0.44 &       0.27 &     2.17 &     2.36 &      0.38 &     0.41 &        4.73 \\
400 CP                 &     1.70 &     2.77 &     0.30 &     0.45 &       0.20 &     2.16 &     2.38 &      0.37 &     0.42 &          NA \\
400 Smooth CP          &     1.71 &     2.77 &     0.30 &     0.45 &       0.20 &     2.16 &     2.38 &      0.37 &     0.43 &          NA \\
400 EasyUQ             &     1.75 &     2.72 &     0.26 & \it 0.44 &   \it 0.12 &     2.16 &     2.10 &  \it 0.35 &     0.39 &    \it 4.33 \\
400 Smooth EasyUQ      &     1.60 & \it 2.61 & \it 0.25 & \it 0.44 &       0.13 & \it 2.15 & \it 2.09 &      0.37 & \it 0.35 &    \it 4.33 \\
\midrule 
2L MC Dropout          & \it 1.45 &     2.19 &     0.33 &     0.41 &       1.07 & \it 1.92 &     1.95 &  \it 0.33 &     0.47 &        4.63 \\
2L Laplace             &     1.64 &     2.29 &     0.22 &     0.44 &         NA &     2.01 &     2.15 &      0.42 &     0.41 &        4.65 \\
2L Single Gaussian     &     1.89 &     2.27 &     0.25 &     0.41 &       0.11 &     2.03 &     2.04 &      0.45 & \it 0.25 &        4.69 \\
2L CP                  &     1.70 &     2.47 &     0.24 &     0.42 &       0.11 &     2.02 &     2.02 &      0.38 &     0.36 &          NA \\
2L Smooth CP           &     1.71 &     2.48 &     0.24 &     0.42 &       0.11 &     2.03 &     2.02 &      0.38 &     0.36 &          NA \\
2L EasyUQ              &     2.07 &     2.40 &     0.24 &     0.42 &   \it 0.03 &     1.98 &     1.83 &      0.42 &     0.30 &    \it 4.30 \\
2L Smooth EasyUQ       &     1.66 & \it 2.14 & \it 0.21 & \it 0.40 &       0.04 &     1.97 & \it 1.82 &      0.40 &     0.27 &        4.31 \\
\bottomrule
\end{tabular}
}
\end{sidewaystable}

To allow comparison with the basic form of EasyUQ, which generates discrete predictive distributions, we use Table \ref{tab:CRPS_ML} and the mean $\CRPS$ from \eqref{eq:CRPS} to assess predictive performance.  Each of the three blocks in the table allows for a seven-way comparison between either Deep Ensembles or MC Dropout, the Laplace approach, the Single Gaussian technique, Conformal Prediction in its basic (CP) and smoothed form (Smooth CP), the basic version of EasyUQ, and Smooth EasyUQ.  As noted, the Naval, Wine, and Year datasets are distinctly discrete, with 51, 6, and 89 unique outcomes, respectively.  For data of this type, predictive distributions ought to be discrete.  Accordingly, there are no benefits of using Smooth EasyUQ for these datasets, as compared to using basic EasyUQ, which adapts readily to discrete outcomes.  In six of the $3 \times 3 = 9$ seven-fold comparisons for the discrete datasets, the basic version of EasyUQ achieves the lowest mean score.  Across the remaining seven datasets and for all three network configurations, smoothing is beneficial, and Smooth EasyUQ outperforms the basic version of EasyUQ.  In 15 of the $3 \times 7 = 21$ seven-fold comparisons on these datasets, Smooth EasyUQ achieves or shares the top score.  All but one of the binary comparisons between Smooth CP and Smooth EasyUQ, all but two of the comparisons between the Single Gaussian technique and Smooth EasyUQ, all but one of the comparisons between the Laplace method and Smooth EasyUQ, and all but eight of the comparisons between Deep Ensembles or MC Dropout and Smooth EasyUQ, are in favour of the latter. 

\section{Discussion}  \label{sec:discussion} 

In this paper we have proposed EasyUQ and Smooth EasyUQ as general methods for the conversion of single-valued computational model output into calibrated predictive distributions, based on a training set of model output--outcome pairs and a natural assumption of isotonicity.  Contrary to recent comments in review articles that lament an ``absence of theory'' (\cite{Abdar2021}, p.~244) for data-driven approaches to uncertainty quantification, the basic version of EasyUQ enjoys strong theoretical support, by sharing optimality and consistency properties of the general Isotonic Distributional Regression (IDR, \cite{Henzi2021a}) method.  The basic EasyUQ approach is fully automated, does not require any implementation choices, and the generated predictive distributions are discrete.  The more elaborate Smooth EasyUQ approach developed in this paper generates predictive distributions with Lebesgue densities, based on a kernel smoothing of the original IDR distributions, while preserving the key properties of the basic approach.  Code for the implementation of IDR in Python \cite{Python} and replication material for this article is openly available \cite{Walz2022}. 

The method is general, handling both discrete outcomes, with the basic technique being tailored to this setting, and continuous outcomes, for which Smooth EasyUQ is the method of choice.  It applies whenever single-valued model output is to be converted into a predictive distribution, covering both the case of point forecasts, as in the WeatherBench example, and computational model output in all facets, such as in the machine learning example, where EasyUQ and Smooth EasyUQ convert single-valued neural network output into predictive distributions.  Percentiles extracted from the predictive distributions can be used to generate prediction intervals. 

The proposed term EasyUQ stems from various desirable properties.  First, the basic version of EasyUQ does not involve tuning parameters nor require user intervention.  Second, EasyUQ operates on the natural, easily interpretable and communicable assumption that larger values of the computational model output yield predictive distributions that are stochastically larger.  Third, EasyUQ is an output-based technique, i.e., it merely requires training data in the form of model output--outcome pairs $(x_i, y_i)$ as in \eqref{eq:data}, without any need to access the potentially high-dimensional covariate or feature vector $z_i$, which serves as input to the computational model that generates $x_i$.  This property is shared with the widely used Single Gaussian technique and related methods, such as the early Geostatistical Output Perturbation (GOP, \cite{Gel2004}) approach and the Quantile Regression Averaging (QRA, \cite{Nowotarski2015}) method for the generation of prediction intervals. 

The term Conformal Prediction (CP, \cite{Marx2022, Vovk2020a, Vovk2005}) refers to a family of output-based methods that yield predictive distributions and prediction intervals that enjoy attractive out-of-sample coverage guarantees, but often entail that shape and scale of the predictive distributions do not vary with the model output.  In simple problems, where predictive distributions that essentially are translates of each other are appropriate, both CP and EasyUQ perform well, and typically yield very similar predictive performance, as illustrated by the temperature example in Section \ref{sec:BasicWB}.  The flexibility of EasyUQ, which allows for predictive distributions that vary in shape and/or scale, subject to the isotonicity condition, materializes in more challenging problems, where predictive distributions that are translates of each other fail.  While EasyUQ adapts to such settings without any need for user intervention, CP might suffer considerable loss in predictive performance, even if adapted manually, as exemplified in the precipitation example in Section \ref{sec:BasicPrecip}.

While adaptive variants of CP are available, their predictive performance in both simulated and real-data settings has been mixed, compared to standard variants \cite{Vovk2020a}.  Recently, Bostr{\"o}m et al.~\cite{Bostrom2021} investigated Mondrian (i.e., covariate-conditional) CP as a flexible alternative, where conformal predictive distributions are built on separate categories formed by binning covariates (in our case, the model output).  This requires additional implementation decisions, namely, on the choice of the bins.  Bostr{\"o}m et al.~\cite{Bostrom2021} take five bins with equal numbers of training instances, which improves predictive performance in their experiments.  From a methodological point of view, in situations where the isotonicity assumption of IDR is met, the binning approach of Mondrian CP can be understood as an approximation to EasyUQ.  EasyUQ finds optimal binnings without manual intervention (\cite{Henzi2021a}, Thm.~2), and training borrows strength from the entirety of training data, whereas Mondrian CP diminishes the training sample by splitting it, which introduces a trade-off between training data size and adaptivity.  A limitation of EasyUQ is that estimates under isotonicity constraints tend to be inconsistent at the boundary of the covariate domain \cite{Guntuboyina2018}, which bears the danger of disproportionately decreased spread of EasyUQ distributions at extreme values of the model output.  In settings where this is of concern, a potential remedy is to resort to Mondrian CP at extreme values, while reaping the benefits of EasyUQ at moderate values of the model output.  We leave further methodological development in these directions to future work.

In contrast to CP and EasyUQ, input-based methods such as MC Dropout \cite{Gal2016}, Deep Ensembles \cite{Lakshminarayanan2017}, the techniques proposed by Camporeale and Care \cite{Camporeale2021} and Chung et al.~\cite{Chung2021}, and the reference methods considered by Duan et al.~\cite{Duan2020} require access to the covariate or feature vector $z_i$.  Input-based methods are much more flexible than output-based methods and thus have higher potential in principle, as evidenced by the success of ensemble methods in numerical weather prediction \cite{Bauer2015, Gneiting2005a}.  However, they tend to be more computationally intense than output-based methods, and as the machine learning example in our paper shows, may not outperform the latter.  Generally, sophisticated input-based methods for uncertainty quantification might realize their potential when applied to substantively informed, highly complex computational models, as in the case of numerical weather prediction, where predictive uncertainty varies.  Output-based approaches to uncertainty quantification typically are less complex and easier to implement, and might nonetheless yield competitive predictive performance when applied to output from data-driven models, such as the neural network models in the benchmark setting from machine learning.

We end the paper with speculations about the usage of EasyUQ and Smooth EasyUQ in weather prediction.  The current approach to forecasts at lead times of hours to weeks rests on ensembles of physics-based numerical models \cite{Bauer2015, Gneiting2005a} but it is being challenged by the advent of purely-data driven models based on ever more sophisticated neural networks \cite{EbertUphoff2023, Schultz2021}.  Published only recently, the WeatherBench comparison \cite{Rasp2020} showed a huge performance gap between forecasts from physics-based numerical models and neural network based, purely data-driven forecasts, with the latter being clearly inferior, as exemplified in our Tables \ref{tab:CRPS_WB} and \ref{tab:LogS_WB}.  Fast breaking developments suggest that the situation may have reversed since, with purely data-driven approaches now outperforming physics-based forecasts of univariate weather quantities \cite{BenBouallegue2023, Bi2022, Chen2023, Lam2022}.  There is a caveat, though, as under the new, data-driven paradigm, spatio-temporal and inter-variable dependence structures might get misrepresented, due to the lack of physical constraints in the model and a need for hierarchical temporal aggregation in the generation of weather scenarios \cite{Bi2022, EbertUphoff2023}.  However, the resulting neural network based forecasts can be subjected to EasyUQ and Smooth EasyUQ, and samples from the resulting predictive distributions can be merged by empirical copula techniques such as ensemble copula coupling (ECC, \cite{Schefzik2013}), to adopt and transfer spatio-temporal and inter-variable dependence structures in physics-based ensemble forecasts.  Hybrid approaches of this type might combine and extract the best from both traditional physics-based and emerging data-driven approaches to weather prediction, and may turn out to be superior to either.  

\section*{Acknowledgements}

We thank Sam Allen, Steffen Betsch, Timo Dimitriadis, Alexander Jordan, Gregor Koehler, Sebastian Lerch\add{, and Jan St\"uhmer} for insightful comments, and we are grateful to Peter Knippertz for many a time discussions, meteorological advice and insights, and encouragement.  Special thanks to Gregor Koehler for suggesting and implementing code for GPU computation.  Two anonymous referees provided a wealth of constructive feedback, for which we are most grateful.  Eva-Maria Walz and Tilmann Gneiting have been working in project C2 of the Transregional Collaborative Research Center SFB/TRR 165 ``Waves to Weather'' funded by the German Science Foundation (DFG).  Alexander Henzi and Johanna Ziegel gratefully acknowledge funding from the Swiss National Science Foundation.  Tilmann Gneiting thanks the Klaus Tschira Foundation for continuous support.

\appendix

\section{Consistency of smoothed conditional CDF estimates}  \label{app:consistency}  

In this appendix, we prove the uniform asymptotic consistency of smoothed estimators of conditional CDFs under mild conditions.  We operate in a very general setting, in which the smoothed estimate, 
\begin{align}  \label{eq:general_check_F}
\check{F}_{x;n}(y) = \int_{-\infty}^\infty \hat{F}_{x;n}(t) \, K_{h_n}(y-t) \, \textrm{d}t 
\end{align}
arises from a basic estimate $\hat{F}_{x;n}$ that uses a sample of the form \eqref{eq:data} of size $n$.  We do not make further assumptions on the form or origin of the basic estimate, though in \eqref{eq:consistency} we specialize to EasyUQ.  As in the main text, we let $K_h(u) = (1/h) \, \kappa(u/h)$ for a smooth probability density $\kappa$, such as a Gaussian or a Student-$t$ density, but we now allow for the possibility that the bandwidth $h_n > 0$ varies with the sample size.  

In formulating the subsequent consistency result, we only require that the basic estimates $\hat{F}_{x;n}$ are asymptotically consistent, and that the true conditional CDFs $F_x(y)$ are smooth in $y$, and we put mild assumptions on $\kappa$.  IDR and its special case EasyUQ indeed are asymptotically consistent under reasonable assumptions.  Specifically, let $(X_{ni}, Y_{ni}) \in \cX \times \real$ for $i = 1, \dots, n$ be a triangular array of covariates and real-valued observations, which are independent across $i$ for any fixed $n = 1, 2, \dots$ and have the same distribution as a pair $(X, Y)$ with conditional CDFs $F_x(y) = \myP(Y \leq y \mid X = x)$.  Let $\hat{F}_{x;n}$ be the IDR CDF computed from this sample with an arbitrary admissible interpolation method for $x \not\in \{ X_{n1}, \dots, X_{nn} \}$.  Here, $\cX$ is some subset of $\real^d$ that is equipped with a partial order $\preceq$.  The key assumption is that the conditional CDFs $F_x$ are nondecreasing in stochastic order, i.e., $x \preceq x'$ implies that $F_x(y) \geq F_{x'}(y)$ for all $y \in \real$.  For continuous covariates, one furthermore needs to assume that the covariate values become sufficiently dense in $\cX$ as $n$ increases, and that a uniform continuity assumption on the conditional CDFs holds; cf.~the references discussed below.  Then the following assumption on uniform consistency is satisfied.

\begin{assumption}  \label{ass:consistent}
There exists a sequence $(\varepsilon_n)_{n = 1, 2, \dots}$ such that, for all $x$ in some set $\cX_n \subseteq \cX$, we have
\begin{align}  \label{eq:hat_F_consistent}
\lim_{n \to \infty} \myP \left( \sup_{y \in \real, \, x \in \cX_n} |\hat{F}_{x;n}(y) - F_x(y)| \geq \varepsilon_n \right) = 0.
\end{align}
\end{assumption}

The sequence of sets $(\mathcal{X}_n)_{n = 1, 2, \dots}$ in the above assumption usually consists of all points in $\mathcal{X}$ whose distance from the boundary of $\mathcal{X}$ is not less than $\delta_n > 0$, where $\delta_n$ is a sequence that converges to zero. Multivariate covariates are treated in Henzi et al.~\cite{Henzi2021a}, who demonstrate Assumption \ref{ass:consistent} with $\varepsilon_n = \varepsilon > 0$ for any constant $\varepsilon > 0$, and $\delta_n = \delta$ for any constant $\delta > 0$.  The case $\cX = (a,b) \subset \real$ with the usual total order corresponds to the typical setting for EasyUQ and is treated by M\"osching and D\"umbgen \cite{Moesching2020}, who show that one can choose $\varepsilon_n$ of order $(\log(n)/n)^{\alpha/(2\alpha + 1)}$ if the conditional CDFs are H\"older continuous in $x$ with index $\alpha \in (0,1]$.  In this case, the sets $\cX_n$ are of the form $(a + \delta_n, b - \delta_n)$ with $\delta_n$ converging to zero at rate $(\log(n)/n)^{1/(2\alpha +1)}$.  The case of ordinal covariates in a finite set was investigated by El Barmi and Mukerjee \cite{ElBarmi2005}, and it can be shown that one can choose $\varepsilon_n = (\log(n)/n)^{1/2}$, and consistency holds for all values of $x$.  While the authors do not explicitly state this convergence rate, it follows from the last displayed equation prior to their Thm.~1, according to which the maximal error (in sup-norm) of the IDR CDFs is less or equal to the error of the empirical CDFs stratified by the covariate.  The sup-norm error of the empirical CDFs can be bounded by $(\log(n)/n)^{1/2}$ (by the Dvoretzky-–Kiefer-–Wolfowitz inequality, $\log(n)$ could be replaced by any other sequence diverging to $\infty$), so the rate stated above applies.

A situation of particular applied relevance arises for distributional single index models (DIMs, \cite{Henzi2021b}), which can be interpreted as a special case of EasyUQ.  Specifically, let $(Z_{ni}, Y_{ni}) \in \zz \times \real$ for $i = 1, \dots, n$, where $n$ is a positive integer, be a triangular array of covariates and observations, which are independent across $i$ for any fixed $n$ and have the same distribution as some pair $(Z, Y)$.  Suppose that there is a function $\theta \colon \zz \to \real$, called the index function, such that the conditional CDFs $F_x(y) = \myP(Y \leq y \mid \theta(Z) = x)$ are stochastically ordered in $x$.  For each sample size $n$, the index function is estimated by $\hat\theta_n$.  Denote by $\hat{F}_{x;n}$ the IDR CDF computed from the pseudo observations $(\hat\theta_n(Z_{n1}),Y_{n1}), \dots, (\hat\theta_n(Z_{nn}),Y_{nn})$ with an arbitrary admissible interpolation method for $x \not\in \{ \hat\theta_n(Z_{n1}), \dots, \hat\theta_n(Z_{nn}) \}$.  If the index function is estimated consistently at a sufficiently fast rate and the pseudo covariates $\theta(Z_i)$ become sufficiently dense in some interval $\cX \subset \real$, then Assumption \ref{ass:consistent} is satisfied with a rate $\varepsilon_n$ of $(\log(n)/n)^{1/6}$, and $\mathcal{X}$ of the form $(a + \delta_n, b - \delta_n)$ for $\delta_n = (\log(n)/n)^{1/6}$.

The second assumption is a natural condition on the true conditional CDFs $F_x$, without which one would not want to smooth in the first place.

\begin{assumption}  \label{ass:Lipschitz} 
There exist constants $L > 0$ and $\alpha \in (0,1]$ such that for all $x \in \cX$ and $u, v \in \real$, 
\begin{align*}
|F_x(u) - F_x(v)| \leq L \, |u-v|^\alpha.
\end{align*}
\end{assumption}

In particular, Assumption \ref{ass:Lipschitz} is satisfied with $\alpha = 1$ if the conditional distributions admit Lebesgue densities that are uniformly bounded.

\begin{theorem}  \label{thm:smoothing} 
Suppose that Assumptions \ref{ass:consistent} and \ref{ass:Lipschitz} hold, and assume that $c_{\kappa,\alpha} = \int_{-\infty}^\infty |s|^\alpha \, \kappa(s) \, \rm{d}s$ is finite.  Then,
\begin{align}  \label{eq:limit}
\lim_{n \to \infty} \myP \left( \, \sup_{y \in \real, \, x \in \cX_n} |\check{F}_{x;n}(y) - F_x(y)| \geq \varepsilon_n + L c_{\kappa,\alpha} h_n^\alpha \right) = 0.
\end{align}
\end{theorem} 

An immediate consequence is that if $h_n^\alpha = \cO(\varepsilon_n)$ then the smoothed estimate $\check{F}_{x;n}$ admits the same convergence rate as the basic estimate $\hat{F}_{x;n}$.

\begin{proof}[Proof of Theorem \ref{thm:smoothing}]
The error of $\check{F}_{n;x}$ is upper bounded as
\begin{align*}
|\check{F}_{x;n}(y) - F_x(y)| 
& = \left\vert \int_{-\infty}^\infty [\hat{F}_{x;n}(t) - F_x(t) + F_x(t) - F_x(y)] \, K_{h_n}(y-t) \, \textrm{d}t  \right\vert \\ 
& \leq \tilde{\varepsilon}_n(x) + \int_{-\infty}^\infty |F_x(y - h_n s) - F_x(y)| \, \kappa(s) \, \textrm{d}s,
\end{align*}
where $\tilde{\varepsilon}_n(x) = \sup_{z \in \real} |\hat{F}_{n;x}(z) - F_x(z)|$, as we see by making the change of variable $s = (y-t)/h_n$.  By Assumption \ref{ass:Lipschitz} we obtain that
\begin{align*}
\int_{-\infty}^\infty |F_x(y - h_n s) - F_x(y)| \, \kappa(s) \, \textrm{d}s \leq L c_{\kappa,\alpha} h_n^\alpha.
\end{align*}
Hence 
\begin{align*}
|\check{F}_{n;x}(y) - F_x(y)| \leq \tilde{\varepsilon}_n(x) + L c_{\kappa,\alpha} h_n^\alpha, 
\end{align*}
and the claim now follows from Assumption \ref{ass:consistent}. 
\end{proof}

We emphasize that Assumption \ref{ass:consistent} is a high-level condition that is not specific to IDR, DIM, or EasyUQ.  Theorem \ref{thm:smoothing} implies that any sequence of conditional CDF estimates $(\hat{F}_{x;n})$ that is consistent in the sense of Assumption \ref{ass:consistent} can be smoothed consistently via \eqref{eq:general_check_F}, either with a fixed kernel $\kappa$, or with a kernel $\kappa$ selected from a suitably limited class of candidate functions.  This is a result of independent interest that goes well beyond the classical setting of smoothing an empirical distribution.  To give an example, the approach can be applied to smoothing conditional CDFs estimators under weaker stochastic dominance constraints \cite{Henzi2021c}, where Thm.~\ref{thm:smoothing} directly yields consistency of the smoothed estimator.  Another method where smoothing might be beneficial is the distributional random forest technique \cite{Cevid2020, Schlosser2019}, which, like IDR, generates discrete estimators of conditional CDFs.  However, the conclusions from Thm.~\ref{thm:smoothing} do not apply directly in this case, since only pointwise but not uniform consistency has been proven for these estimators (\cite{Cevid2020}, Corollary 5).

\bibliographystyle{abbrv}
\bibliography{refs}

\end{document}